\documentclass[twocolumn,floatfix,superscriptaddress,longbibliography,nofootinbib]{revtex4-1}
\RequirePackage[sort&compress]{natbib}

\usepackage{color}

\usepackage{float}

\usepackage{amssymb,amsfonts,amsmath}
\usepackage{bm}
\usepackage[pdftex]{graphicx}
\usepackage{xcolor}
\usepackage{comment}

\usepackage[colorlinks = true,
            linkcolor = blue,
            urlcolor  = blue,
            citecolor = blue,
            anchorcolor = blue]{hyperref}

\usepackage{dcolumn}
\usepackage{bm}

\usepackage{xcolor}

\renewcommand{\[}{\left[}
\renewcommand{\]}{\right]}

\usepackage{xcolor}
\definecolor{RoyalBlue}{HTML}{4169e1}
\definecolor{ForestGreen}{HTML}{228b22}

\usepackage{setspace}
\usepackage{comment}

\usepackage{todonotes}

\newcommand{\rev}[1]{{\color{black} #1}}

\begin{document}


\title{Emergence of complex network topologies from \rev{flow}-weighted optimization of network efficiency}

\author{Sebastiano Bontorin}
\affiliation{Fondazione Bruno Kessler, Via Sommarive 18, 38123 Povo (TN), Italy}
\affiliation{Department of Physics, University of Trento, Via Sommarive 14, 38123 Povo (TN), Italy}

\author{Giulia Cencetti}
\affiliation{Fondazione Bruno Kessler, Via Sommarive 18, 38123 Povo (TN), Italy}

\author{Riccardo Gallotti}
\affiliation{Fondazione Bruno Kessler, Via Sommarive 18, 38123 Povo (TN), Italy}

\author{Bruno Lepri}
\affiliation{Fondazione Bruno Kessler, Via Sommarive 18, 38123 Povo (TN), Italy}

\author{Manlio De Domenico}
\email[Corresponding author:~]{manlio.dedomenico@unipd.it}%
\affiliation{University of Padua, Via Francesco Marzolo 8, 35131, Padua, Italy}
\affiliation{Padua Center for Network Medicine, University of Padua}

\date{\today}

\begin{abstract}
Transportation and distribution networks are a class of spatial networks that have been of interest in recent years. These networks are often characterized by the presence of complex structures such as \rev{central} loops paired with \rev{peripheral} branches, which can appear both in natural and man-made systems, such as subway and railway networks.
In this study, we investigate the conditions for the emergence of these non-trivial topological structures in the context of human transportation in cities.
We propose a minimal model for spatial networks generation, where a network lattice acts as a spatial substrate and edge velocities and distances define an effective temporal distance which quantifies the efficiency in exploring the \rev{urban space.}
Complex network topologies \rev{can be} recovered from the optimization of joint network \rev{paths} and we study how the interplay between a \rev{flow} probability between two nodes in space and the associated travel cost influences the resulting optimal network. 
In the perspective of urban transportation we simulate these flows by means of human mobility models to obtain Origin-Destination matrices.
We find that \rev{when using} simple \rev{lattices}, the \rev{obtained} optimal topologies transition from tree-like structures to more regular networks, depending on the spatial range of \rev{flows}. 
\rev{Remarkably}, we find that branches paired to large loops structures appear as optimal structures when the network is optimized for an interplay between heterogeneous mobility patterns of small range travels and longer range ones typical of commuting. Finally, we \rev{show that our framework is able} to recover the statistical spatial properties of the \rev{Greater London Area} subway network.

\end{abstract}

\maketitle

\section{Introduction}

Cities represent one of the most fascinating man-made complex systems, exhibiting complex features ranging on different scales: from their structure and dynamical behavior, up to the scaling of socio-economic factors with their size \cite{Batty_2008, Barthelemy_2019, Bettencourt2010, Pan2013, Arcaute2015}. These features represent a strong hint towards the existence of universal underlying mechanics behind apparently very different cities \cite{Bettencourt2013, Bassolas2019, Bettencourt2020}. Out of these structural properties, one of the most relevant, as it plays a fundamental role mediating the complex interplay \rev{between} human dynamics \cite{Schlpfer2021, Barbosa2018} and mobility in urban context, are transportation networks \cite{Alessandretti2022, Lee2017, Gallotti2021, Barthelemy_2011, Morris2012}.
These networks are a class of spatial networks whose properties have been investigated in the literature during the \rev{last two decades} \cite{Barthelemy_2011, Gastner_2006}. In particular, they have been studied under the lens of optimality conditions and minimization of cost-based functionals \cite{Gastner_2006}, in order to identify specific features behind efficient networks. The concept of optimal networks \cite{Barthelemy_2019} and energy-like minimization \cite{Barthelemy2006} has its natural understanding in the physics language. States of a system which minimize a functional defining trade-offs between system's observables (e.g., free energy) represent the most likely to be observed states of many real world systems. While in some complex systems, such as cities, these physical variables can not be derived from first principles, these analogies and concepts can still offer a valid perspective and provide an embedding of these systems in a space where the interplay between their structure and dynamics can be unfolded and better understood. Simple laws have been studied \cite{Louf_2013, Gastner_2006} to better understand the emergence of hierarchy and the role of traffic in the network state. Moreover, global and local optimization criteria lie in the evolution of man-made systems where policy makers and planners can adopt some of these criteria in their plans \cite{Barthelemy_2011}.\\

Transportation networks are often characterized by the presence of complex structures \cite{Banavar1999, Pei_2022, Gallotti2015} such as loops paired with branches \cite{Roth2012}, which can appear both in natural \cite{Tero2010} and man-made systems \cite{Barthelemy_2011}, \rev{like} railway and subway networks. These structures represent the key topological elements behind efficient public transportation systems \cite{Pei_2022}.
In this study we investigate the conditions for the emergence of these non-trivial structures \cite{Louf_2014, Louf_2013} in the context of human transportation in cities. We aim to reconstruct these topologies by means of an optimal configuration \cite{Ibrahim2021} of the network state. Under the assumption of a fixed total cost and a limited set of high-capacity connections (e.g., a constraint in the expenditure available on infrastructure), the optimal configuration is the assignation of connections' velocities, or edges' weights, such that the joint amount of time required to travel between two nodes is minimized for all pairs of nodes. Moreover, as these networks represent the arteries in urban exploration/navigation via public transportation, we study the role of traffic \cite{Zhang2015, Louf_2013} in weighting the importance of specific set of connected edges (paths). We model the urban morphological structures which generate heterogeneous distributions of human mobility in space, biasing these optimal networks to converge towards specific topological features.
We aim to explore the minimum requirements and the conditions for these optimization processes to reproduce the empirical structures aforementioned.\\

At variance with the recent works on network efficiency, we adopt some fundamentally different modeling choices.
We evaluate the efficiency in terms of time necessary to explore the network, where edges' weights act as \rev{travel} speeds. We also weight path efficiency by the traffic probability between nodes. The underlying network lattice (as represented in its simplest form by the triangular lattice in the next sections) acts as a substrate that allows the network to evolve \cite{Szell2022} and possibly exhibit the network topological features typical to real world systems.
On this framework, we show how \rev{introducing} simple probabilities biasing the optimal efficiency between points in space force a transition between a tree-like topology and a network resembling a simple lattice. We show also that the modeling of traffic-like \rev{flows} forces the emergence of preferential shared paths in space. The optimal configuration of these shared paths leads to complex topologies, which ultimately shows features seen in real systems. Features such as a bi-modality in the edges' velocity distribution, characteristic of multi-layered transportation, and the central core with loops paired with branches typical of subway systems \cite{Roth2012, Pei_2022} are recovered. We finally show an application of the model within the \rev{Greater London Area}, finding similarities of the optimal model with its \rev{London Underground} network.

\begin{figure}
    \centering
    \includegraphics[width=0.5\textwidth]{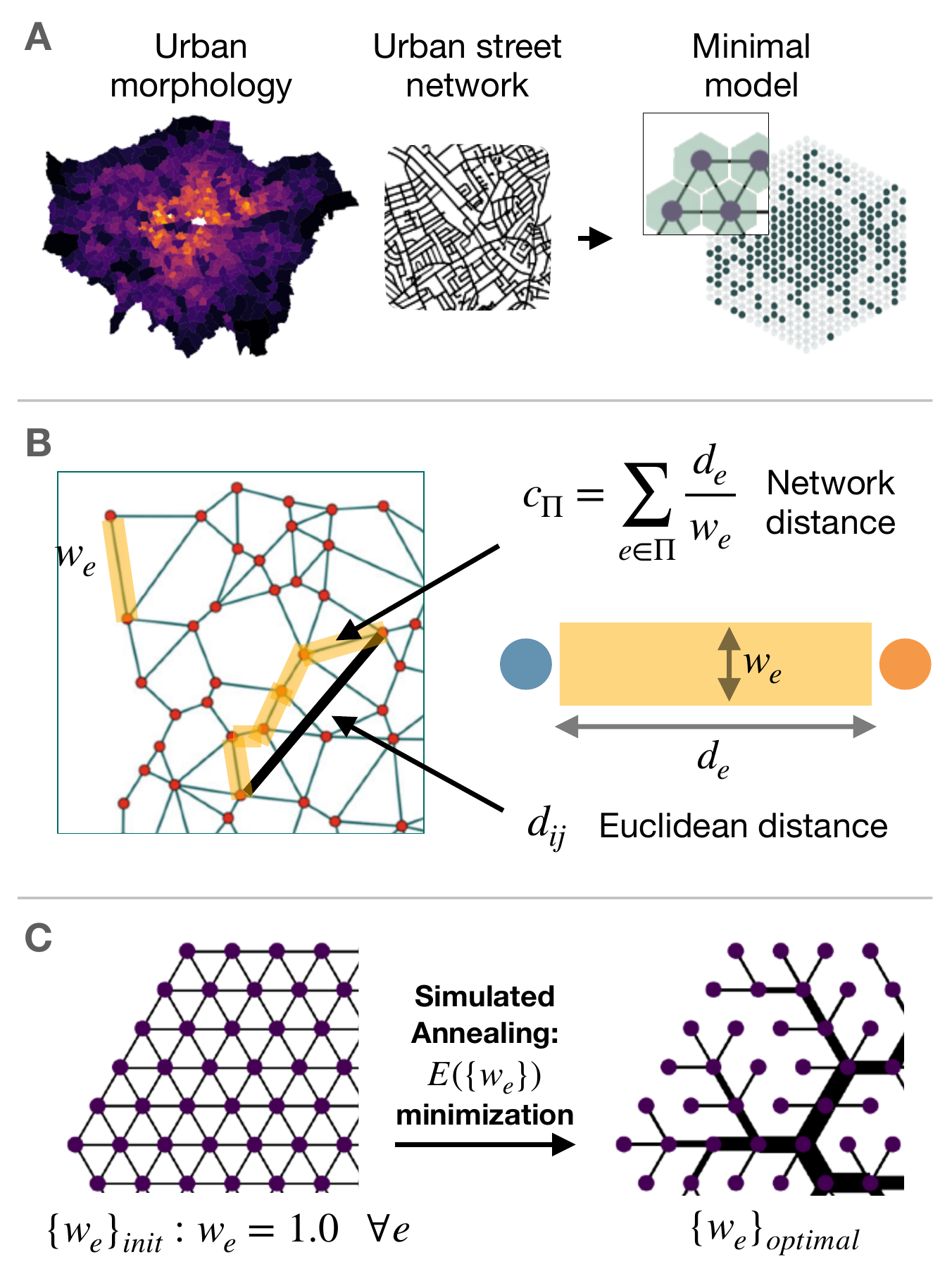}
    \caption{ \small{ \textbf{Spatial network model for urban morphology}: \textbf{A)} Mapping population distribution and urban transportation network to a minimal spatial network where nodes encode urban features. Example with hexagonal tiling mapped to the triangular lattice. \textbf{B)} Network-based distance $c_{\Pi_{ij}}$ versus euclidean distance $d_{ij}$; edges weights/velocity $w_{e}$ are depicted as widths.  \textbf{C)} Edges weights of the lattice substrate are optimized via simulated annealing to unravel spatial features of the optimal transportation network.
   }}
   \label{fig:model_scheme}
\end{figure}

\section{Framework for Urban Spatial Morphology}

We introduce here a general framework for spatial networks with the aim of recovering a minimal model for urban morphologies that encode both transportation properties and urban features such as population and density of points of interest (POIs). To this aim, we begin from the definition of a network substrate which acts as an effective discretization of the spatial dimension. Its simplest form can be found in an hexagonal 2-dimensional tiling \cite{Birch2007} and its planar dual, the triangular lattice. More formally, in this network substrate each tile in space is represented by a node, connected to its set of neighboring nodes (see Fig. \ref{fig:model_scheme}). The existence of a physical edge between nodes/tiles $i$ and $j$ is encoded in the adjacency matrix $A$ where $A_{ij} = 1$ if the regions are neighbors in the lattice. Distances and metrics are therefore computed on top of this network substrate and are biased by nodes' features.

\begin{figure*}[!]
    \centering
    \includegraphics[width=\textwidth]{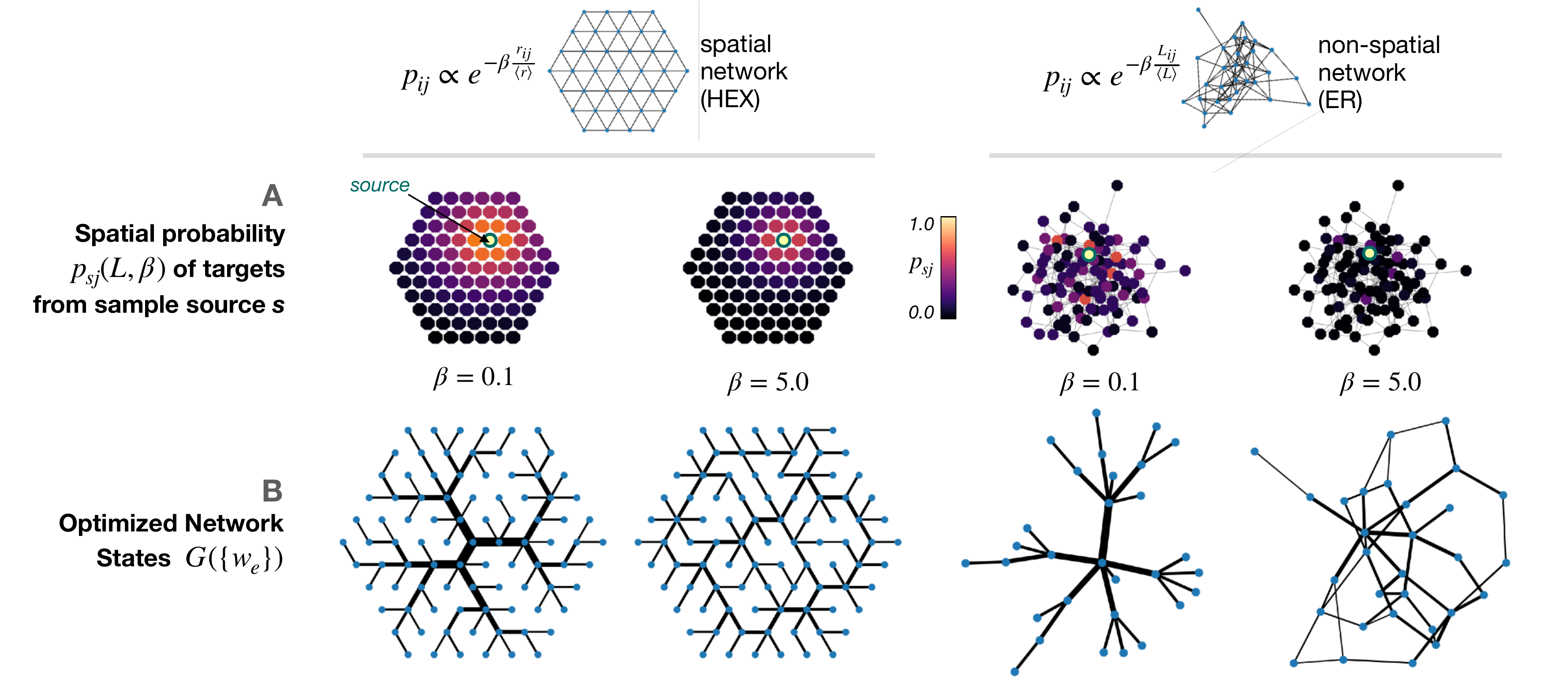}
    \caption{ \small{ \textbf{Optimization of synthetic networks}: The role of $\beta$ is studied for two \rev{network} models: the triangular lattice (HEX) and the non-spatial (ER) network. \textbf{A)} Heatmap of target nodes probabilities $p_{ij}$ from source node (yellow) under two different $\beta$ values: as the penalty parameter grows, farther nodes are more penalized and \rev{flows} tend to stay close to the source. \textbf{B)} Samples of the associated optimized network states: when \rev{flows} are not affected by distances ($\beta$ = 0.1) source nodes target all the other nodes in the network with approximately equal probability, the optimal network converges to a tree-like structure. With larger $\beta$ ($\beta$ = 5.0), trip probabilities are more localized and the presence of loops appear in the optimal structure.
   }}
   \label{fig:simple_models}
\end{figure*}

Nodes of this network \rev{can} encode spatial features at the urban scale, such as population or amenities' density in a given node.
We therefore have a minimal representation of a urban morphological structure (see Fig. \ref{fig:model_scheme}), and a network substrate that acts as a transportation system and can be optimized to generate optimal transportation networks \cite{Szell2022}.
The path-based temporal distance on top of the transportation network acts as the fundamental metric we aim to minimize. The rationale behind a network-based distance is grounded on the assumption that in the context of public transportation, urban systems are not navigated by considering geographical distance but rather by \rev{evaluating the} travel-time between \rev{departure and arrival}. More specifically, multi-layer transportation networks \cite{Gallotti2015, Alessandretti2022} are characterized by layers having a hierarchical organization with different characteristic speeds \cite{Gallotti2016}. Thus, an effective temporal distance becomes fundamental in determining accessibility and efficiency in urban space exploration.\\

\begin{figure}
    \centering
    \includegraphics[width=0.5\textwidth]{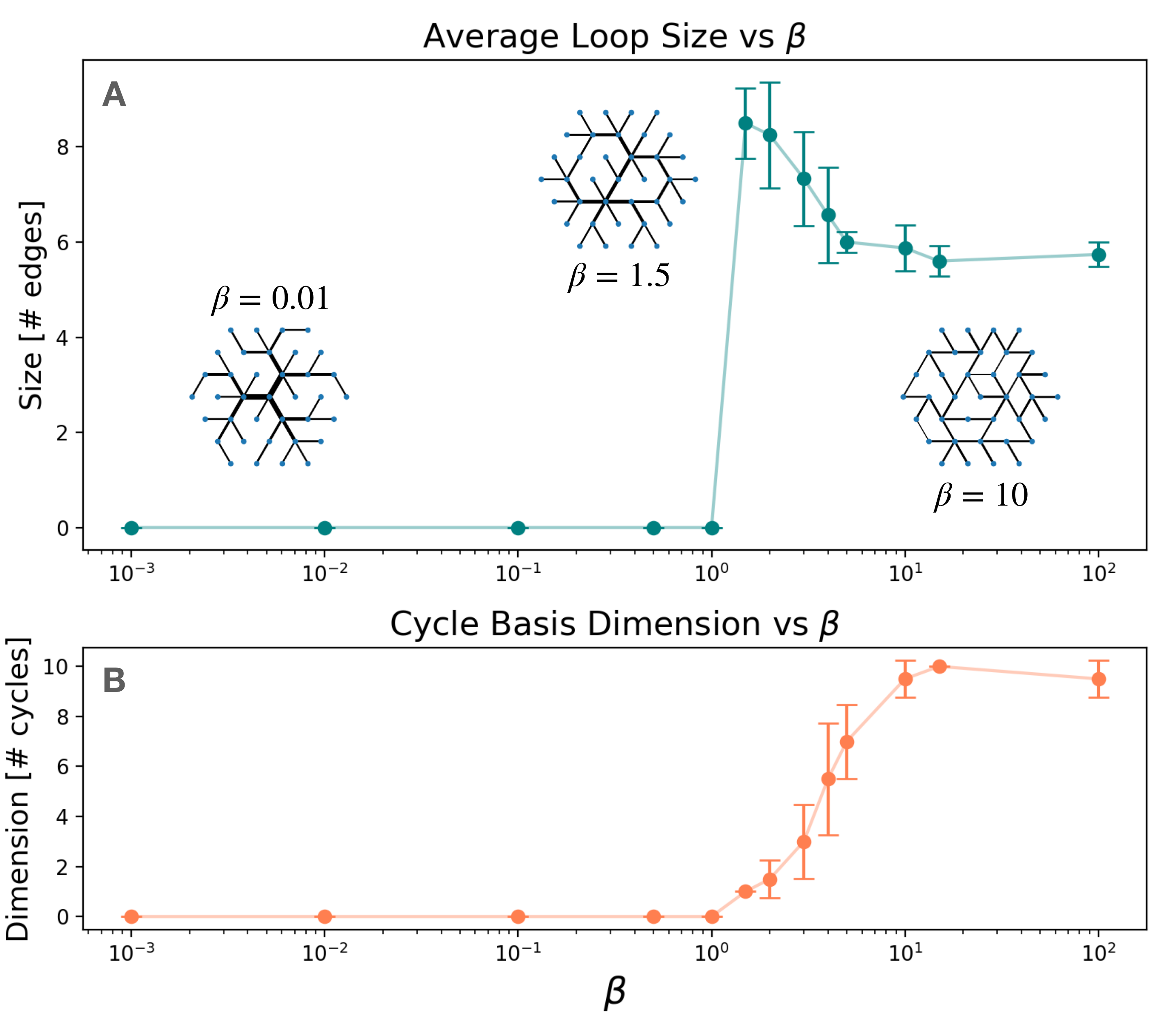}
    \caption{ \small{ \textbf{Loop Dimension vs $\beta$}: Minimum cycle basis is used as a network's observable to study the appearance of loops. For each point the median and its absolute deviation are shown. \textbf{A)} The average size (number of edges) of the loops that constitute the cycle basis. \textbf{B)} Cycle basis dimension as number of loops. The optimal network ranges from a tree structure to a lattice-like with small loops, as the probability of long range movement decreases (large $\beta$).
    A transition in the cycle basis property is observed at \rev{$\beta \sim 1.0$} for the triangular lattice under study, where the optimal network results in an intermediate state with large loops.
   }}
   \label{fig:cyclebasis}
\end{figure}

In this model, we denote $e$ as an $edge$ in the network, and $w_{e}$ as the \rev{associated edge's weight} which can be seen as a \rev{velocity} of the edge in the transportation network. $d_{e}$ is the euclidean distance of edge $e$ between the nodes it is connecting; here edge weights are visually mapped as widths of the links. Information about edges distance in this framework can be relevant when generalizing to the case of random spatial networks where edges have different lengths. In the case of a general non-spatial network, where there is no notion of spatial distances, the model can be adapted by fixing $d_{e}= 1$. Finally, we define $\Omega_{\Gamma_{ij}}$ as the set of paths connecting the two nodes. We then maximize the efficiency of this underlying substrate. The transportation efficiency between two nodes $i-j$ is computed as a cost in terms of time \cite{Ren_2014}, \rev{and we do not account for congestion, which can be introduced in a possible extension of this framework}. We find the path (a set of connected edges starting from source node $i$ and ending in destination node $j$) with the smallest cumulative time, where the time delay introduced by choosing en edge $e$ is measured as a ratio between \rev{its euclidean distance  and its weight, representing a proxy of speed}. 
Refer to Fig. \ref{fig:model_scheme} for a graphic depiction.
Here $G(\{w_{e}\})$ is used to indicate the network configuration with the associated set of edges weights $\{w_{e}\}$.
We therefore aim to find the assignment of weights $\{w_{e}\}$ as a trade-off between network efficiency in transportation, by minimizing the set of costs $\{ c_{ij} \}$ of travelling between pair of nodes $i-j$ where each element $c_{ij}$ is defined as:

\begin{equation}
    c_{ij}(\{ w_{e}\}) = \min_{\Pi \in \Omega_{\Gamma_{ij}}} \left[\sum_{e \in \Pi_{ij}} \frac{d_{e}}{w_{e}} \right],
    \label{path-cost}
\end{equation}

\rev{and in} absence of further information, the optimization procedure is the equivalent of minimizing the travel costs $\sum_{ij} c_{ij}$.
Here we add a novel ingredient, in which we couple the optimization of the network temporal distances with a traffic \rev{flow} or probability between pairs of nodes. Operationally, when dealing with real world Origin-Destination (OD) matrices, this probability can be then mapped to a traffic $T_{ij}$ between two points. $T_{ij}$ represents the probability of a person from node $i$ to travel to node $j$, and a traffic can be recovered when information about populations in source and target nodes is added.
$T_{ij}$ effectively acts as a rank in the importance of a specific path in the network. As paths connecting different pairs of nodes may share common edges of the network substrate, complex topologies emerge from the shared paths jointly optimizing the network efficiency. The \rev{flow}-weighted transportation efficiency therefore becomes:


\begin{equation}
    E(G(\{w_{e}\})) = \frac{1}{N(N-1)} \sum_{i}^{N} \sum_{j \neq i}^{N} T_{ij} \cdot c_{ij}(\{w_{e}\})
    \label{eq:efficiency}
\end{equation}

We also require that the total network \rev{infrastructure} cost, defined as the cumulative sum of edges weights per unit length, multiplied by edge distance $C_{G} = \sum_{e \in G} d_{e}w_{e}$ is conserved.
This is a generalization of a standard optimization process, \rev{in the sense that when $T_{ij} = 1$, $\forall (i,j)$, the efficiency is optimized
for all possible trip pairs $(i,j)$ with equal importance}, where the Minimum Spanning Tree often represents the optimal solution \cite{Barthelemy_2011}.

\rev{Before tackling the problem of traffic-like (OD) \rev{flows}, we study a simpler definition of $T_{ij}$,} which allows to understand the role of \rev{distance in the optimization process, in absence of other nodes' features}:


\rev{
\begin{equation}
    T_{ij} \propto e^{-\beta d_{ij}}.
    \label{eq:distance-dependence}
\end{equation}

The coefficient $\beta$ appearing in Eq. \ref{eq:distance-dependence} is introduced as a penalizing parameter and determines how relevant is the pair-wise distance $d_{ij}$ when computing probabilities. We can understand it as the inverse of a characteristic traveling distance for an agent on the network $\beta \sim \frac{1}{d_{0}}$. 
While several alternatives on the integration of distance in spatial-dependent probabilities (such as power-laws $T_{ij}  \sim d_{ij}^{-\gamma}$) can be employed, we focus on the exponential dependence as it represents the foundational result from the maximum entropy derivation of gravity \rev{flows} \cite{Wilson_1975}. The introduction of gravity-like \rev{flows} will be discussed in Section \ref{sec:attractiveness}.}

 
In the following, we introduce the application of the model on simple substrates to explore \rev{the role of $\beta$} in absence of spatial urban features.


\rev{\section{Optimization of simple network substrates}}

In order to asses the role of the characteristic distance parameter $\beta$ in the emergence of specific topologies, we compute networks statistics on a set of generative models for both spatial and non spatial networks.
As hexagonal tiling of space is preferable when an analysis includes aspects of connectivity \cite{Birch2007}, the first model we study is a triangular lattice. The \rev{reason} behind this choice is that it represents the planar dual  \cite{Barthelemy_2011, Viana2013} of the hexagonal lattice. Therefore, as space is discretized in hexagonal tiles, the spatial network connecting its centers is the triangular lattice, \rev{which is isotropic and presents less equivalent degenerate paths of a rectangular lattice.} As a direct reference to hexagonal tiling, we refer to this network as HEX (see Fig. \ref{fig:simple_models}). We also extend the analysis also to the case of a random network model where nodes are not embedded in a metric space. Specifically, we study an Erd\H{o}s-Rényi (ER) network topology, where the definition of distance between nodes $L_{ij}$ can be defined in terms of topological shortest path distance \cite{Newman2010}.


\begin{figure*}
    \centering
    \includegraphics[width=0.9\textwidth]{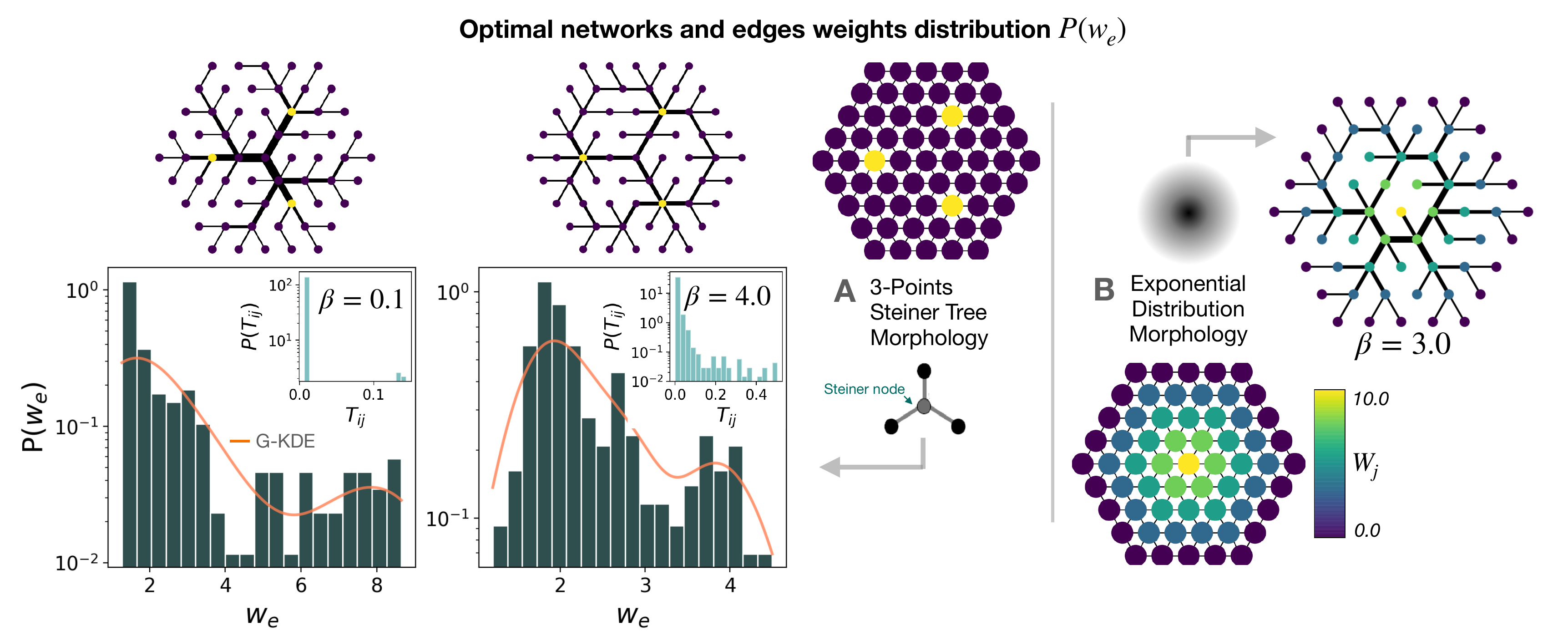}
    \caption{ \small{ \textbf{Minimal models of urban morphology and attractiveness distributions under study (3-Points and Exponential decay):} Morphology of POIs, where attractiveness $W_{j}$ is mapped with color intensity (yellow being higher). Optimized edges weights distributions $P(\{w_{e}\})$ are characterized by the bi-modal nature that reveals the multi-layered structure of the optimal transportation networks when close-range flows are paired with long-range traffic typical of commuting towards city centers (Insets $P(T_{ij})$ with peaks on large-flows due to POIs). Gaussian KDE is shown in orange as a visual aid. \textbf{A}) 3-points polycentric distribution of POIs, \rev{resembling the euclidean Steiner Tree problem \cite{Dreyfus1971, Brazil2013} for three points. The network is} optimized with $\beta = 0.1$ and $\beta = 4.0$, \rev{and} shows the appearance of branches connecting the POIs paired to large loops in the periphery. \textbf{B}) Optimal state and distribution of speeds with exponential decay of $w_{j}$ from the center and an exemplifying result with $\beta = 3.0$. The optimal topology is characterized by a central loop paired with branches.
   }}
   \label{fig:simple_attractiveness}
\end{figure*}

As a first benchmark we \rev{simplify \rev{flows} as a spatial probability $T_{ij} = p_{ij}$ that decays exponentially with distance and does not consider nodes features}, the resulting equation for \rev{$p_{ij}$ is}:

\begin{equation}
  p_{ij} = \frac{ e^{ - \beta d_{ij}/ \langle d \rangle}}{\sum_{k \neq i}^{N-1} e^{ - \beta d_{ik}/ \langle d \rangle}}  
\label{eq:simple-probabilities}
\end{equation}

Where $\langle d \rangle = \frac{1}{N (N-1)}\sum_{i \neq j} d_{ij}$ is the average distance of points in the network and acts as a normalization factor (euclidean distance $\langle r \rangle$ in case of a spatial network or topological $\langle L \rangle$ for the ER network). 

Therefore $p_{ij}$ encodes how much of the nearby space is explored by a single source node. An illustration of the spatial dependence of target probabilities and samples of the resulting optimal topologies are presented in Fig. \ref{fig:simple_models}.

For a range of $\beta$ values the optimization process is performed on an ensemble of these models. To assess the emergence of complex structures we observe the number of loops that emerge in the optimal state. This measure is relevant in the context of spatial networks, where loops break the symmetry introduced by optimal structures such as trees.
We compute the minimum cycle basis set as a metric to observe the emergence of loops \cite{Kavitha2007}: i.e. the minimum set of loops (\rev{where a single loop is encoded} in a set of edges that defines a \rev{closed path} in the graph) such that any other closed path in the network can be reconstructed via combination of this cycle basis \cite{Kavitha2007}. 
Specifically, we investigate the cycle basis dimension (the number of loops that constitute this set) and the average loop size, against a range of $\beta$ values. \rev{This metric allows to quantify the emergence of spatial topological features that differentiate the optimal state from a tree structure.}
Results for these synthetic systems are presented in Fig. \ref{fig:cyclebasis}. Additional boxplots are shown in \rev{SM Figures 1-2}.  A tree-like topology is recovered when the \rev{flows} \rev{probabilities} are distributed uniformly across all nodes in space (when $\beta \rightarrow 0$ and  distance is \rev{therefore} not a penalizing variable in Eq. \ref{eq:simple-probabilities}), while loops emerge when farther targets become less likely to be explored and the network is globally optimized for close-range trips. \rev{Notably, in Fig. \ref{fig:cyclebasis}} \rev{around $\beta \approx 1.0$}, we observe a sharp transition in the average loop size \rev{in the HEX lattice under analysis}:
connections appear between neighboring nodes which are far from the tree-root as it becomes more efficient to have a direct link.
\rev{In this $\beta$ regime the tree topology does not guarantee the most efficient configuration for peripheral nodes, which have their high probability targets in their direct neighborhood (see Eq. \ref{eq:simple-probabilities}). Thus in the optimization process edges appear between leaves nodes which are in separated branches: this ultimately breaks the tree structure and leads to the emergence of large-scale loops.} Eventually the optimal network converges to a simpler structure with small loops as the network is optimized for nodes to target only direct neighbors in the lattice. Finally, in \rev{SM Section 2} we show an application on the case of a single target node in the perimeter of the lattice, where the model reproduces leaves venation patterns \cite{Barthelemy_2011, Katifori2010}.

\section{Spatial attractiveness and traffic-like flows}\label{sec:attractiveness}

In the context of urban systems, optimal transportation networks need to be devised to accommodate traffic \rev{flows} \cite{Zhang2015} towards specific areas of interest, e.g. due to high commercial and business land use density.
Hence we extend the efficiency optimization framework in the case where we have more realistic traffic on top of the urban networks, as the presence of nodes with high attractiveness (POIs) biases the \rev{flows} towards them. In urban scenarios we adopt spatial-interaction models to mimic more traffic-like \rev{flows}.
\rev{In these models, flows are obtained} via a gravity-like equation: \rev{$T_{ij} \propto p_{i}p_{j}\exp{(-\beta d_{ij})}$} \cite{Barbosa2018} which can be derived from first principles via entropy maximization, thus representing the most likely set of \rev{flows} to be observed.
In the context of urban exploration, the gravity equation can be mapped to a model for spatial interaction \cite{Piovani2017, Wilson_1975} where nodes with a given attractiveness $W_{j}$ compete as possible targets for traffic:

\begin{equation}
T_{i j} \propto \frac{1}{Z} P_{i} W_{j} \exp \left(-\beta d_{i j}\right)
\label{eq:real_traffic}
\end{equation}

Normalization $Z$ accounts for all possible trip alternatives ${\sum_{k} W_{k} \exp \left(-\beta d_{i k}\right)}$. $P_{i}$ is the population density in node $i$ and $W_{j}$ encodes a suitable definition of benefit/attractiveness of node $j$ as a possible target \cite{Piovani2017}. $T_{i j}$ is therefore the fraction of population in node $i$ commuting/travelling on average to node $j$. To better understand the role of nodes' attractiveness, we start with the simplest assumption of equal population distribution on all nodes: $P_{i} = 1.0~\forall i$; we will introduce more realistic population distribution in the next section with the London case study.\\

\begin{figure*}
    \centering
    \includegraphics[width=\textwidth]{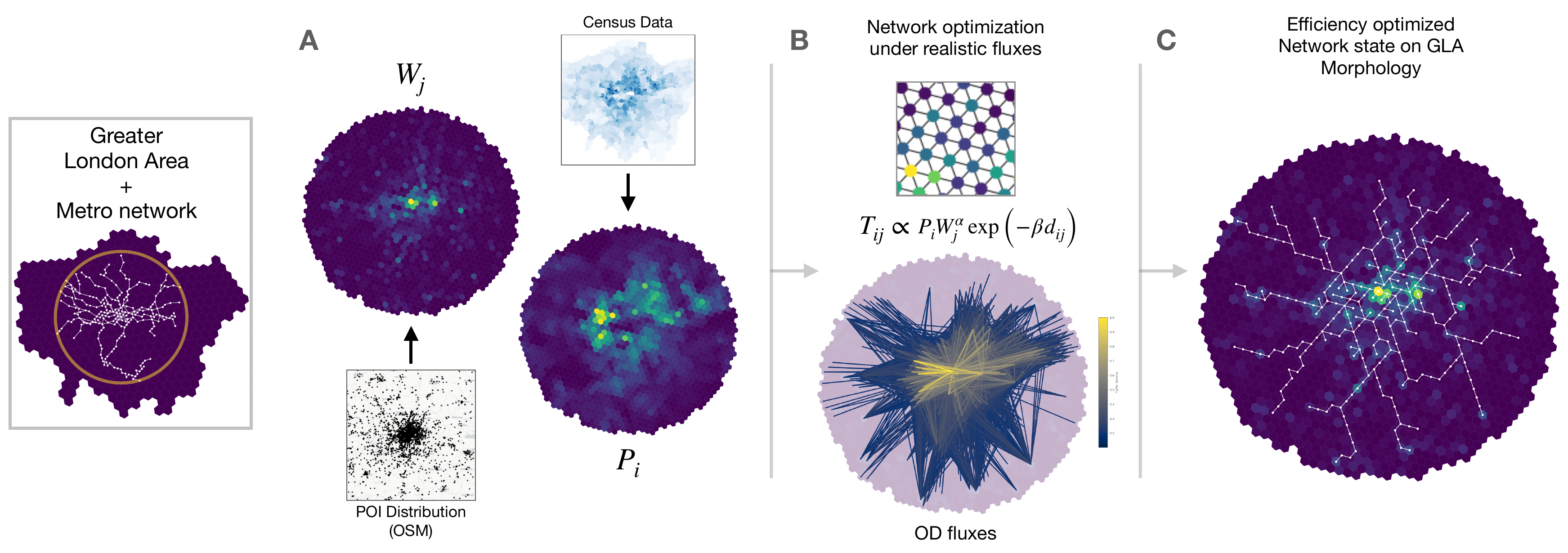}
    \caption{ \small{ \textbf{Optimal network model for \rev{Greater London Area} subway system}: Application of the efficiency optimization with realistic \rev{flows} on the urban structure of the \rev{Greater London Area}. \textbf{A)} Urban morphology data is recovered from Census and OSM and population and POI densities are mapped to the H3 tiling. \textbf{B)} Data is mapped to the triangular lattice, with nodes having features which allow the calculation of traffic-like \rev{flows}, a sample OD matrix is shown where $T_{ij}$ are computed with $\beta = 0.35$. \textbf{C)} Optimal network state for the London model, where only edges and nodes corresponding to the second mode are shown (\rev{see SM Section 3}). Central core structure with loops paired with peripheral branches can be visually seen.
   }}
   \label{fig:LGA-model}
\end{figure*}

We apply these models on the triangular lattice to unravel the optimal topologies that emerge \rev{when traffic probabilities are biased towards some nodes having high attractiveness (simulating POIs)} and we study two spatial configurations for nodes' \rev{$W_{j}$}. In the first configuration high $W_{j}$ is assigned to three nodes (POIs) placed at the vertices of an equilateral triangle. \rev{We study the 3-points distribution as it mimics a prototypical polycentric distribution of city-centers, and it can be linked to the solution of the euclidean Steiner Tree problem \cite{Dreyfus1971, Brazil2013}.}
\rev{The Steiner Tree is a class of problems where given a set of N points in a plane the goal is to find the set of lines connecting the points with minimum cumulative length. In our case, the solution would lie in the central node of the lattice being the Fermat point \cite{Brazil2013} and the Steiner node, which connects the three vertices of the high $W_{j}$ triangle, as illustrated in Fig. \ref{fig:simple_attractiveness} panel A.}
\rev{The second case} is a distribution of $W_{j}$ that decays exponentially from the center, mimicking a more realistic morphology for a urban monocentric structure. The two morphologies are depicted in Fig. \ref{fig:simple_attractiveness}.


We find that due to nodes in the network biasing the traffic \rev{flows}, as it can be seen in the insets \rev{of Fig. \ref{fig:simple_attractiveness} A}, the traffic \rev{flows} get divided into two types: a close range paired to a long range \rev{set of flows}, due to POI polarization. We show \rev{in Fig. \ref{fig:simple_attractiveness}} optimal solutions for \rev{values of} $\beta = 0.1$, $4.0$. \rev{Interestingly, optimal solutions are characterized by three central lines branching from the center (which therefore acts as Steiner node) and connecting the three nodes with high attractiveness, therefore resembling the solution of the Steiner tree problem}. Moreover, in the case of more localized \rev{flows} ($\beta = 4.0$) these \rev{lines} are also paired with large scale loops connecting farther nodes.
We also find that the heterogeneity of traffic \rev{flows} forces the appearance of a second mode in the distribution of speeds $w_{e}$ (see Fig. \ref{fig:simple_attractiveness}).
The two peaks in the optimal $P(w_{e})$ can be interpreted as two different levels of speed, which suggests that the entire process can be decomposed in two distinct mechanisms which can be mapped as a bi-layer network: one layer at high capacity with long-range/commuting trajectories and the other one at low velocity with short-range paths. These two layers can be ideally separated, \rev{hinting towards a possible extension of the model to multilayer networks.}



\section{Greater London Area: generative model for the subway system}

We extend in this section the application of the model by integrating data from a real urban structure. Specifically, we model the morphology of \rev{Greater London Area (GLA)} on top of our framework and apply the efficiency optimization process with the aim of understanding if the temporal efficiency optimization of the spatial substrate paired with realistic flows is sufficient to yield a transportation network with similar topological features (such as a central core paired with peripheral branches \cite{Roth2012}) as the London subway system.
To extend the model to real urban scenarios, we first obtain the distribution of amenities \cite{Hidalgo_2020} from OpenStreetMap \cite{OpenStreetMap} and we use this density of points in space as a proxy to estimate the attractiveness $W_{j}$ of a tile. Census data for Greater London Area yards from 2014 is used to recover population density $P_{i}$. These densities are then mapped to Uber's H3 tiling to recover the spatial discretization in hexagonal tiles, such that we can have direct mapping to the nodes on a triangular lattice, as in the examples discussed in previous sections. We thus have the ingredients to finally simulate the spatial interaction \rev{flows} $T_{ij}$ in Eq. \ref{eq:real_traffic}.
In \rev{Fig. \ref{fig:LGA-model}} the integration of urban data \rev{describing the} London's morphology \rev{in the model} is explained and we provide a depiction of the OD \rev{matrix} that arises from the spatial interaction model. With the aim of reproducing real features, we impose an upper limit on edge weight, so that the distribution of weights is bounded during the optimization process: $w_{e} \in (0, w^{*})$. This better simulates the upper bound in speed of real multilayer systems. 
Further explanation of data recovery and integration in the model is provided in \rev{SM Section 3}. We find (\rev{see SM Figure 6}) that $\{w_{e}\}$ distribution \rev{displays} a bi-modal shape, and this allows the analysis of the generated network in a sub-graph defined by the set of high speed edges. 
In Fig. \ref{fig:LGA-model}, panel C, we show a sample result for $\beta = 0.35$ of this sub-graph. The characterization of the network into a central core paired with peripheral branches as the optimal state can be visually observed.

\begin{figure}[H]
    \centering
    \includegraphics[width=0.5\textwidth]{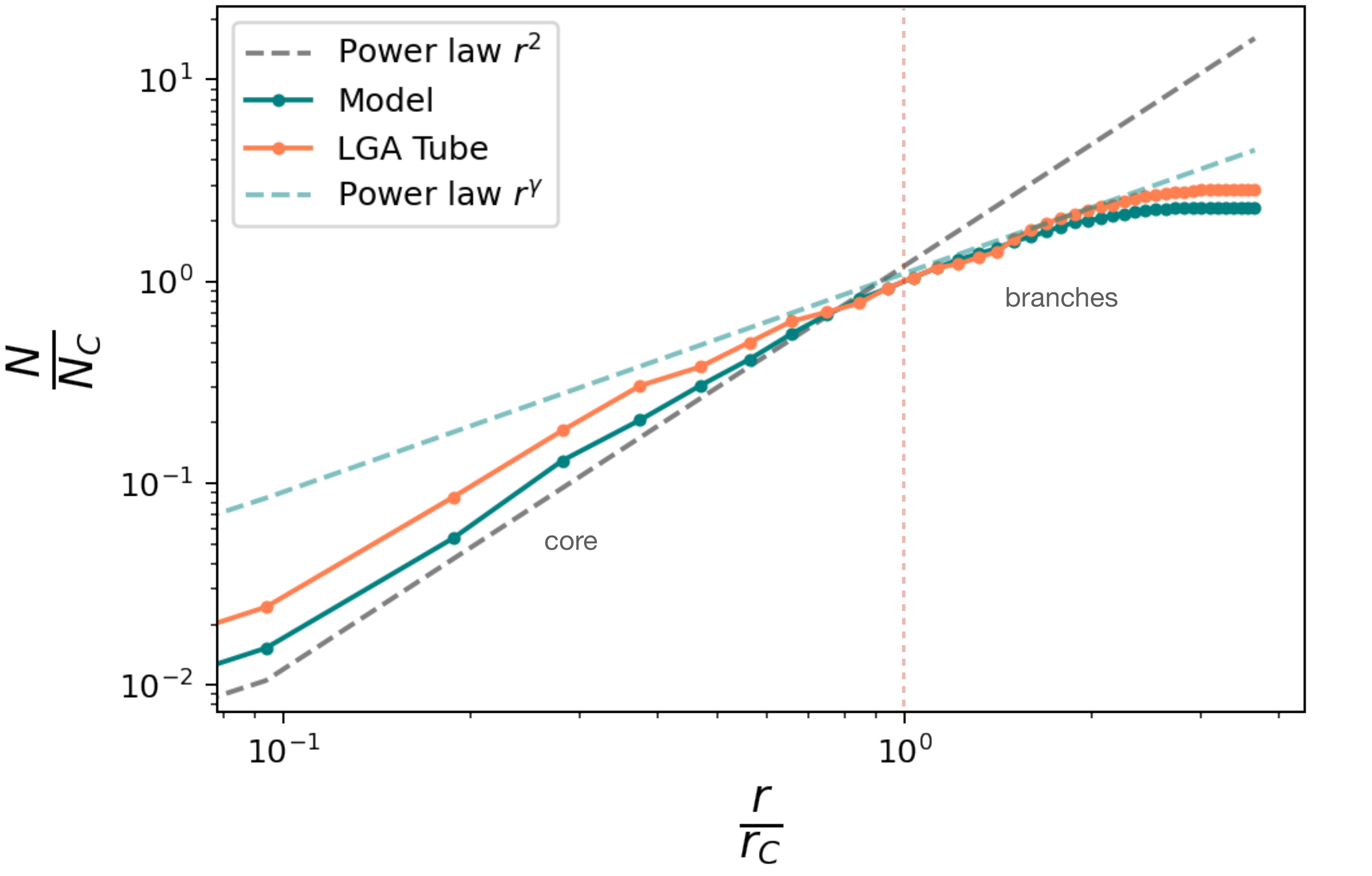}
    \caption{ \small{ \textbf{Scaling properties of GLA Tube stations}: Profile of the  number of stations (nodes in the optimal discretized network (see \rev{SM Section 3}) reproducing GLA underground) versus the distance from the barycenter. The scaling of $N(r)$ profile of the model is compared with the real network system. Scaling properties predicted in \cite{Roth2012} are verified, finding the two different scaling regimes separated at $\frac{r}{r_{C}} \sim 1$ for core paired with branches systems, where $r_{C}$ is the core radius \rev{(characterized by $r^{2}$ scaling)}, and $N_{C}$ is the number of stations in the core. \rev{The scaling exponent $\gamma = 1.25 \pm 0.02$ is obtained as a linear fit of the integral curve \cite{Roth2012} for $r>r_{C}$ (see SM Section 4 for more details).}
   }}
   \label{fig:scaling}
\end{figure}

The model's subgraph of high speed edges is compared to the real tube network in the \rev{Greater London Area} \cite{Gallotti2015} to assess the similarities between the optimal structure and the real subway system. We quantify this similarity by means of spatial scaling laws \cite{Roth2012}, these are convenient to highlight the recovery of the central core structure characterized by loops, paired with quasi mono-dimensional lines branching from the core.
We investigate the distribution of nodes stations using \rev{the} profile function $N(r)$ that quantifies the total number of stations at a distance $r$ from the network barycenter, \rev{computed as the average location of all station nodes \cite{Roth2012}}. Results of this scaling \rev{analysis} for the real and simulated networks are presented in Fig. \ref{fig:scaling}. The two scaling regimes indicate the separation of core and branches: the scaling of $r^{2}$ in the core center and a second trend due to mono-dimensional branches for $r > r_{C}$, where $r_{C}$ is the radius of the core structure. \rev{The} second \rev{trend can be computed analytically via an integral curve for $N(r > r_{C})$ which can be approximated by a power law $r^{\gamma}$ ($\gamma = 1.25 \pm 0.02$, see SM Section 4), as in Ref. \cite{Roth2012}}. The curve of $N(r)$ is consistent with the real network and confirms scaling laws prediction from \cite{Roth2012}.

\section{Discussion}

Starting from simple conditions on temporal efficiency on a spatial network substrate, we show that network optimization paired with traffic-like \rev{flows} weighting the importance of specific connections in space can reproduce complex networks features from man-made transportation networks.
Specifically, we devise a framework for spatial networks where nodes can encode features of urban systems and can ultimately lead to the study of optimal topologies in real scenarios. A key novelty lies in the optimization process happening on a spatial substrate, such that edges of the resulting optimal network are optimized to improve the efficiency of the shared space by all nodes in the network. 
We show how the probabilities of moving from one point to another in space force a transition between a tree-like and a lattice-like topology in the optimal network. Fixing certain target points in space with a higher attractiveness for \rev{flows} can reproduce theoretical results such as the Steiner tree solution or leaves venation patterns. We also show that extending these probabilities using urban spatial information and traffic-like flows modeling forces the emergence of shared preferential paths that are organized as complex topologies, resulting from traffic weighted optimization of network time efficiency, which ultimately exhibits the characteristics seen in real systems. We recover features such as a bi-modality in the speed distribution of the edges of the network, characteristic of multilayer transportation. Or the appearance of a central core with loops coupled to branches typical of underground systems, as in the case of the London underground system. We find that branches paired to large loops structures
appear as optimal structures when the network is optimized for an interplay of traffic \rev{flows} mixed between small range travels and longer range ones typical of commuting.
This novel framework for the optimization of spatial networks in urban contexts may show further improvements and extensions to better accommodate the concepts of multi-layer and shared space. It could be extended also to the case of inter-cities transportation, where \rev{specific} nodes \rev{in the network substrate} represent cities. To conclude, in this work the problem is addressed in a theoretical way with the aim of reproducing and understanding some features observed in real spatial networks, but future works can exploit this framework as a basis to understand how to generate optimal transportation networks in a urban planning scenario.





\subsection*{Competing Financial Interests}
The authors declare no competing financial interests

\subsection*{Data Availability}
The data used in this work are publicly available from the original references

\subsection*{Code Availability}
The code to perform the analysis will be available upon request.

%

\bibliographystyle{apsrev4-1}
\bibliography{biblio}


\clearpage
\onecolumngrid

\setcounter{figure}{0}
\setcounter{equation}{0}
\setcounter{table}{0}
\setcounter{section}{0}

\renewcommand{\thesection}{S\arabic{section}}
\renewcommand{\thesubsection}{S\arabic{section}.\arabic{subsection}}
\renewcommand{\thetable}{S\arabic{table}}
\renewcommand{\theequation}{S\arabic{equation}}

\renewcommand{\figurename}{\textbf{Supplementary Figure S}}
\makeatletter
\def\fnum@figure{\figurename\thefigure}
\makeatother

\section*{Supplementary Material for ``Emergence of complex networks topologies from flow-weighted optimization of network efficiency"}

\section{\label{sec:simple_models} Application On Simple Network Substrates}

In this Section we report more detailed results for the optimization of simple network substrates that was discussed in the Main text, specifically for the triangular (HEX) lattice (planar dual of the hexagonal tiling of space) and Erd\H{o}s-Rényi (ER) network. The aim is to explore the resulting topologies that emerge both in spatial and non spatial networks when simple probabilities are taken into consideration (see Main text).

\subsection{HEX Lattice and ER Network}

Results of model application on the triangular (HEX) lattice and results of model application on an Erd\H{o}s-Rényi (ER) non-spatial network. For the non-spatial generative model, an ensemble of 20 networks with N = 30 nodes and edge probability $\rho = 0.2$ is generated. For the spatial case, the optimization process is repeated 20 times for each value of $\beta$ on the triangular lattice with N = 37 nodes.\\

SM Fig. 1 and SM Fig. 2 show boxplots for the distributions of metrics computed on the cycle basis in panels A on both figures. In panels B, samples of the optimized network states are shown for different values of $\beta$.

\begin{figure}[H]
    \centering
    \includegraphics[width=0.85\linewidth]{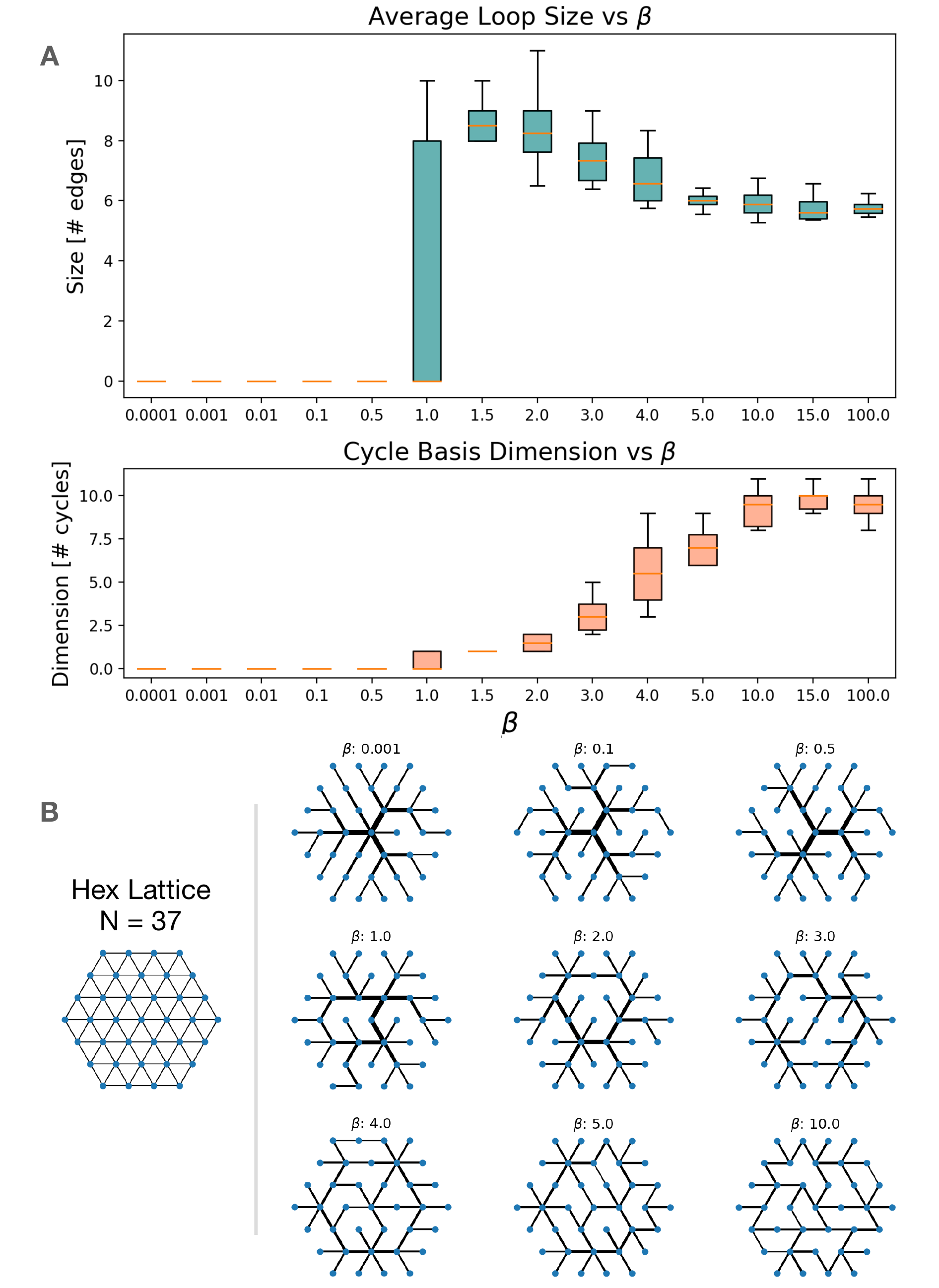}
    \caption{
    \small{
        \textbf{Cycle basis properties and samples - HEX lattice.}
        Boxplot statistics for the cycle basis dimension and optimal network samples for triangular lattice across different $\beta$ values. 
    }
    }
    \label{fig:HexSimple_COMP}
\end{figure}



\begin{figure}[H]
    \centering
    \includegraphics[width=0.85\linewidth]{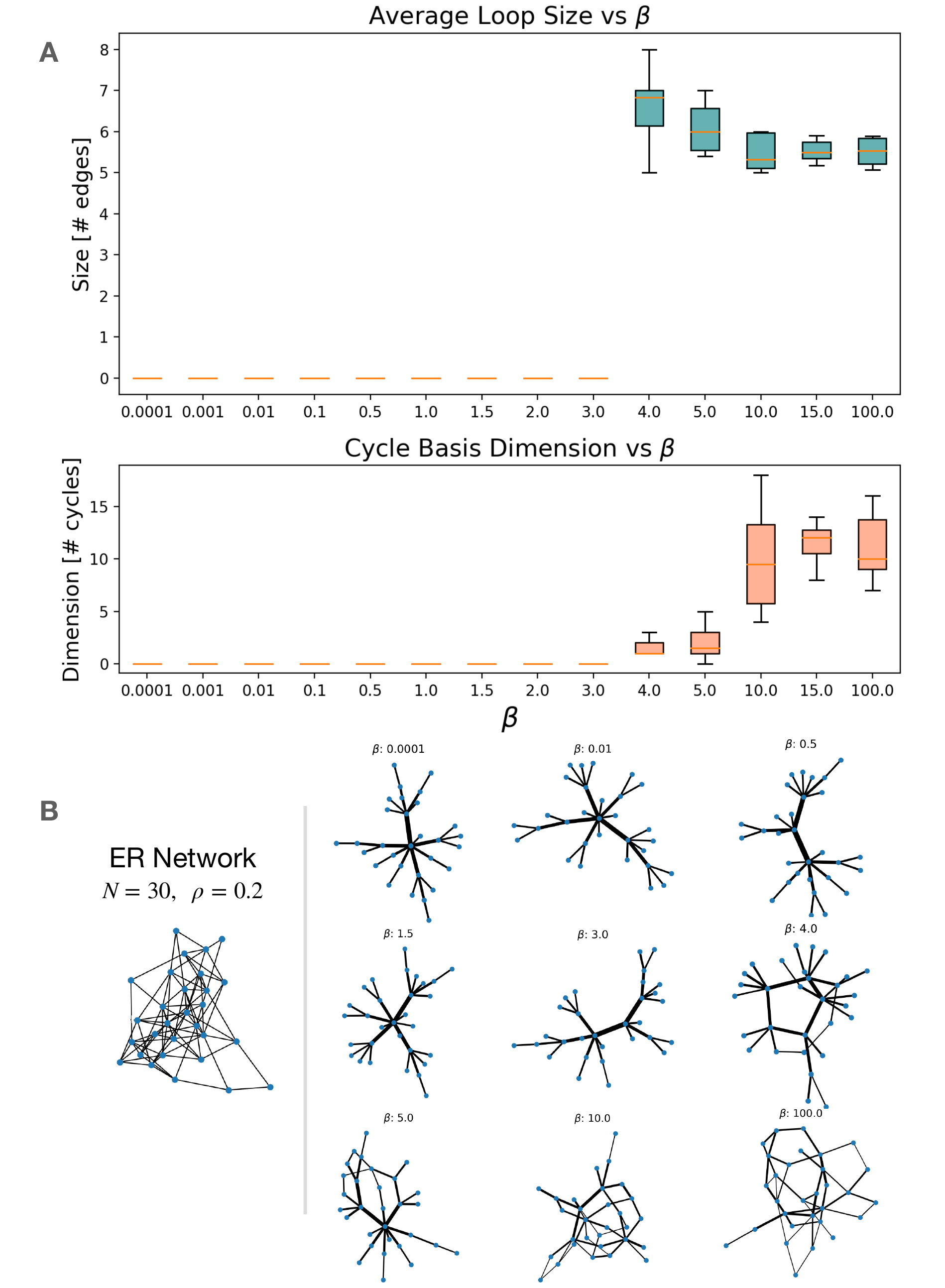}
    \caption{
    \small{
        \textbf{Cycle basis properties and samples - ER network.}
        Boxplot statistics for the cycle basis dimension and optimal network samples for an Erdos-Renyi (ER) network across different $\beta$ values.
    }
    }
    \label{fig:ERSimple_COMP}
\end{figure}

\newpage

\section{Leaves patterns}

Here we show an application of the model to reproduce leaf venation patterns \cite{Katifori2010}. A single attracting node (single target or sink) is considered at one of the perimeter nodes of the lattice, and the substrate is optimized using all the other nodes as sources.

\begin{figure}[H]
    \centering
    \includegraphics[width=0.8\linewidth]{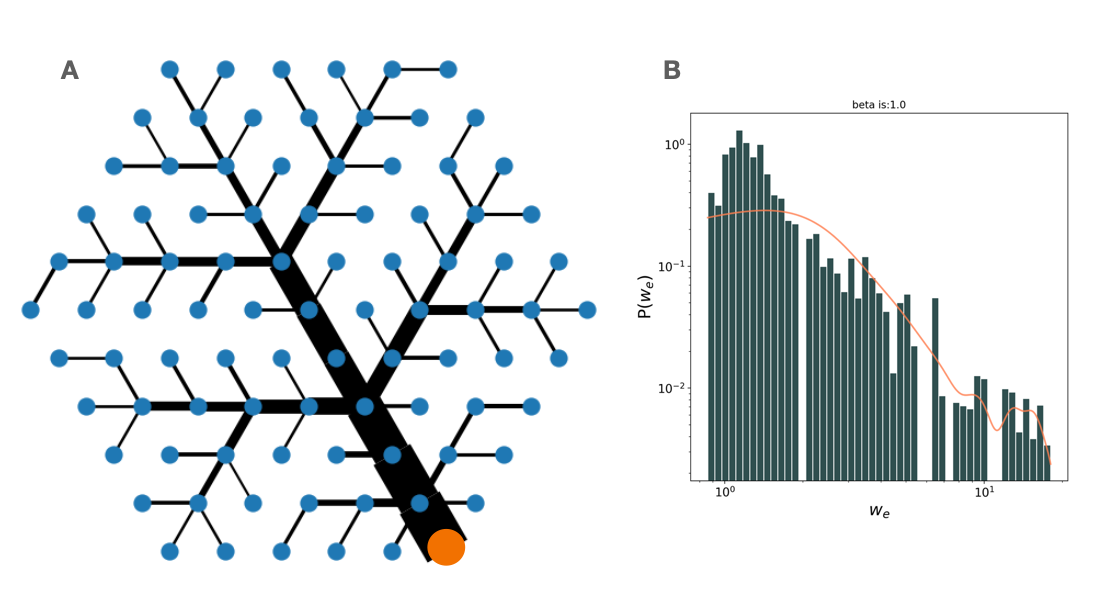}
    \caption{
        \textbf{Optimal networks resembling leaves' veins patterns.}
        Optimal state when a single target (orange node) in a single spatial extremity is considered. Efficiency is optimized for all nodes in space to reach the target. The resulting optimal state resembles tree-like patterns found in leaves (A), while the distribution of edges weights is shown in (B).
    }
    \label{fig:leaves}
\end{figure}

\newpage

\section{Application on London Tube network}

In this section we discuss more in detail the application of the model on the Greater London Area urban morphology. We show that the London subway network spatial properties can be recovered by means of the optimal network state.\\
In particular, to simulate a real transportation system, an upper bound on edges travel velocity is imposed.


\subsection{Data Integration: Census and OpenStreetMap Data}

To extend the model to real urban scenarios, we gather data regarding the urban morphological structure from OpenStreetMap (OSM) \cite{OpenStreetMap} and Census.

\subsubsection{OSM Data}

To model the attractiveness of nodes in the lattice, we use the density of Points Of Interest in the urban space.
We use amenities \cite{Hidalgo_2020} points in OSM as a proxy for Points Of Interest (POIs), and a node $j$ attractiveness ($W_{j}$) therefore encodes the density of amenities in space.

The bounding box for Greater London Area (GLA) is obtained via OSM (\url{https://wiki.openstreetmap.org/wiki/Bounding_Box}) and POI densities are recovered inside this box. Specifically we use the following amenities sub-categories:

\begin{center}
    \texttt{'cafe','college','library','school','university','kindergarten','restaurant','pub',
    'fast food','bar','bank','dentist','pharmacy','hospital','clinic','doctor','arts centre','cinema','community centre','police','post office','marketplace'}
\end{center}

\begin{figure}[H]
    \centering
    \includegraphics[width=0.8\linewidth]{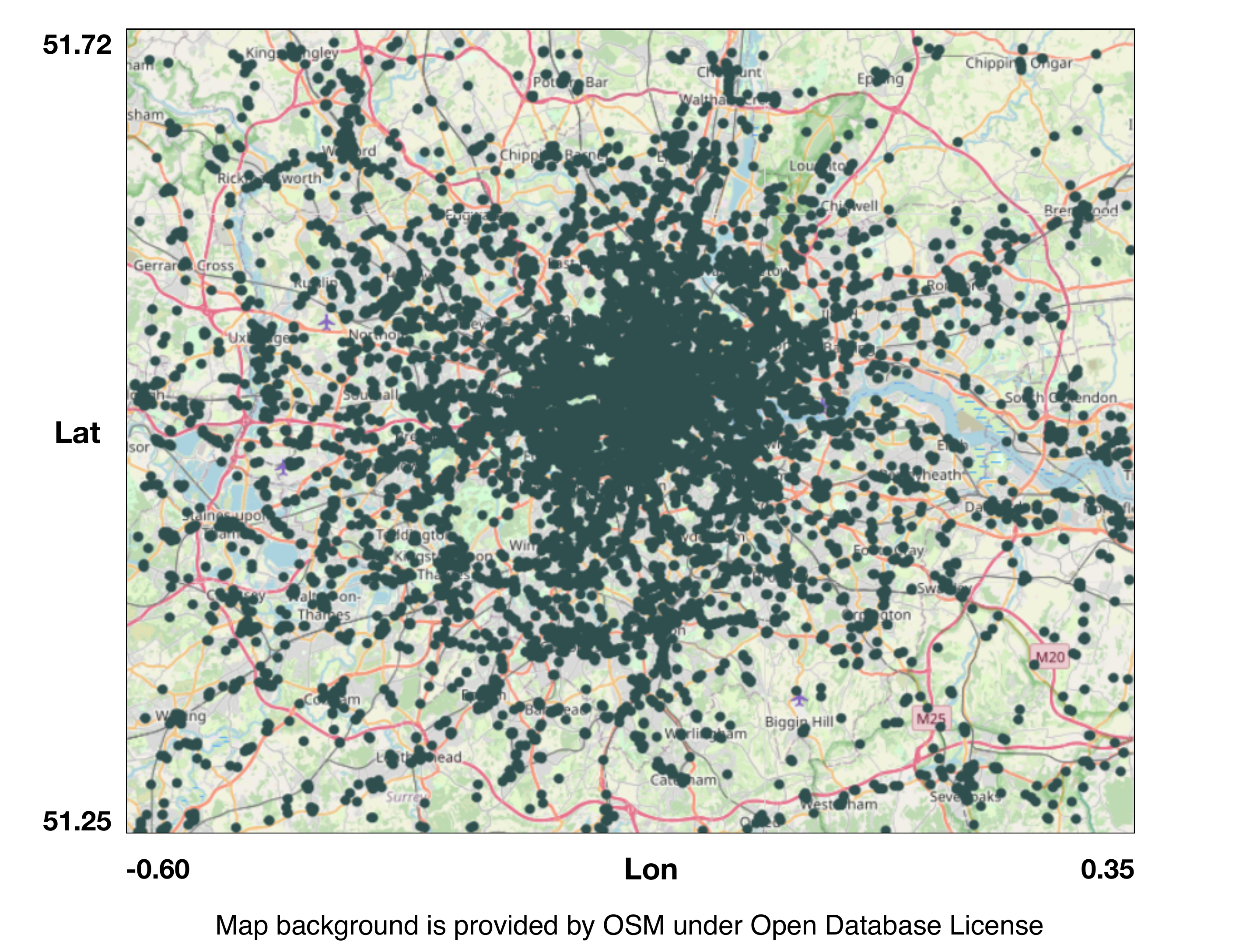}
    \caption{
    \small{
        \textbf{OSM amenities query for GLA bounding box.}
        Amenities retrieved from OSM (\url{https://openstreetmap.org/copyright}) for the bounding box defined by the following longitude $\[-0.6,0.35 \]$ and latitude interval $\[51.25,51.75\]$.
    }}
    \label{fig:OSM}
\end{figure}

In Fig. S\ref{fig:OSM} the amenities points recovered from OSM are plotted in the bounding box.

\subsubsection{Census Data}

\begin{figure}[H]
    \centering
    \includegraphics[width=0.8\linewidth]{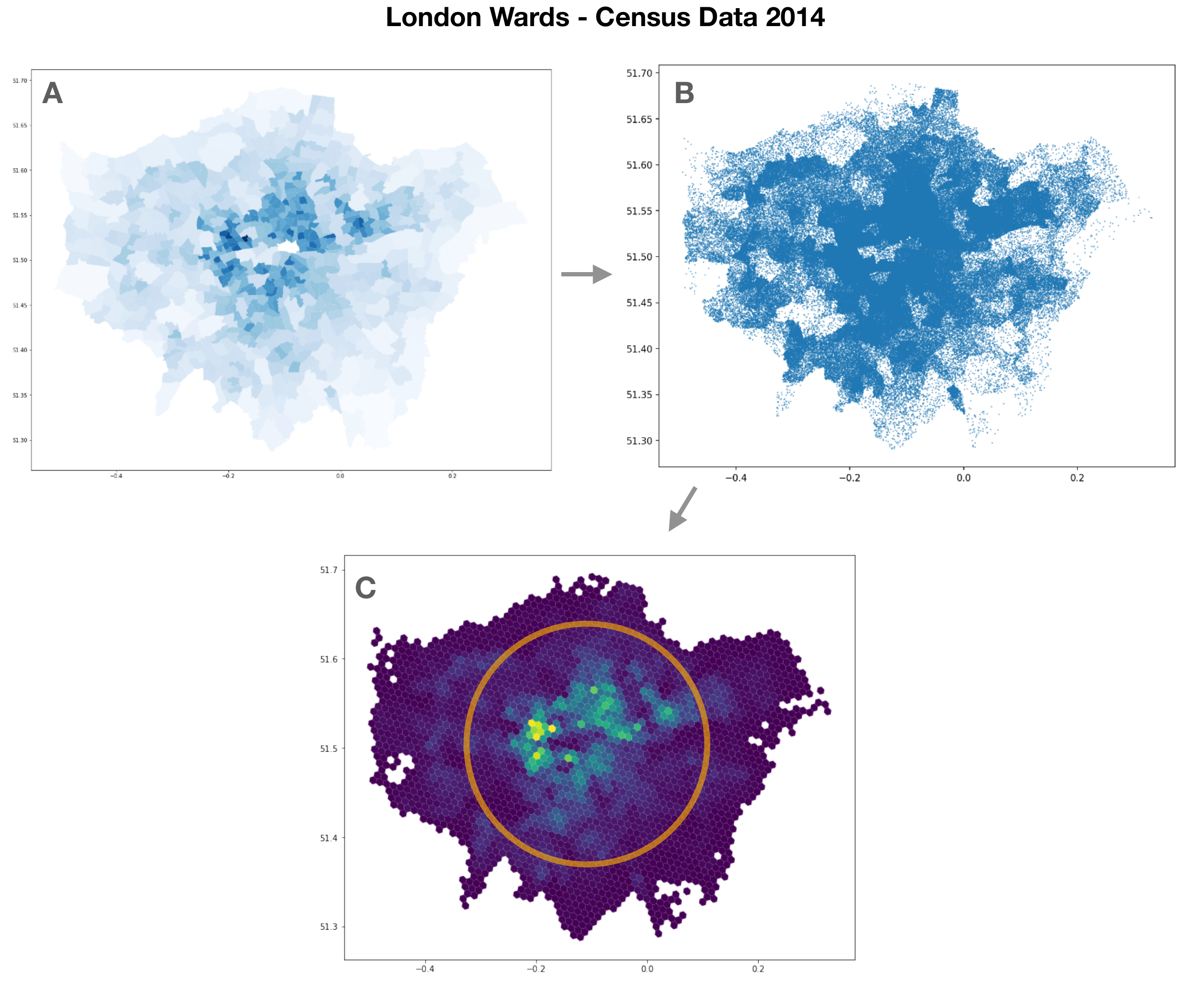}
    \caption{
    \small{
        \textbf{Census data retrieval.}
        \textbf{A)} London's wards data from Census 2014 (\url{https://data.london.gov.uk/dataset/ward-profiles-and-atlas}). \textbf{B)} Points are generated in space following wards polygons with a density proportional to Census data. \textbf{C)} Points are mapped to HEX3 tiles with associated densities of points. Finally, a restriction to a disc is used to enforce symmetry in the distribution of nodes and to ease the computational load.
    }}
    \label{fig:census}
\end{figure}

Points are mapped to density in space by using HEX3 tiling (\url{https://eng.uber.com/h3/}) with spatial resolution $RES$ = 8. This spatial discretization process via tiling is of particular relevance to map this information to the HEX triangular lattice model which was described in the Main text. Both Fig. S\ref{fig:census} and Fig. S\ref{fig:OSM} show how model information $P_{i}$ and $W_{j}$ are recovered from data and then mapped to tiles as in shown in panel C in Fig. S\ref{fig:census}. Tiles covering the urban area are then mapped to the spatial network model as nodes in the triangular lattice (see Fig. 5 in Main text).\\

For computational simplicity and to preserve the isotropy of the lattice substrate from its central point, we restrict our analysis to a disc centered in the London's region with highest attractiveness, that lies approximately in the City Of London district. Results are robust against integration of the remaining GLA region. As the optimal edges weights are influenced by the traffic on the substrate, the discarded regions in the Greater London Area do not add relevant contributions when compared to more central regions with higher population density.


\subsection{Results with distributions for traffic and bounded weights distribution}

To simulate a real system we limit upper edges weights to a fixed value $w^{*} = 7*w_{init}$ where $w_{init}$ is the initial edges weight assigned to the network state ($w_{init} = 1.0$) before optimization process, such that $\sum_{e \in G} d_{e} w_{e, init} = C$. In this simulation we work with $\beta = 0.35$. 


\begin{figure}[H]
    \centering
    \includegraphics[width=0.9\linewidth]{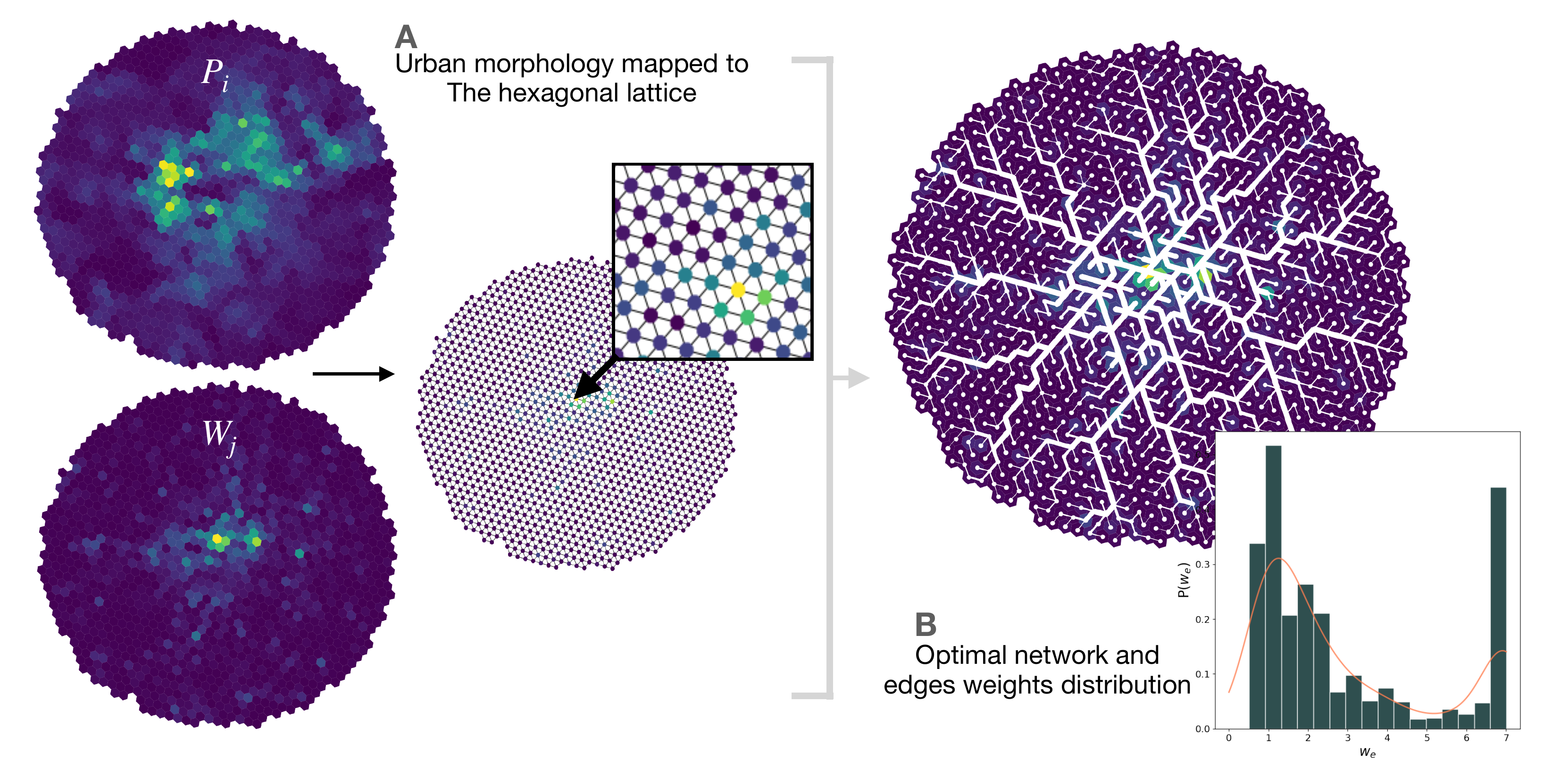}
    \caption{
    \small{
        \textbf{Mapping the data to Hex lattice + GLA optimized network.}
    \textbf{A)} Urban morphology data mapped to the HEX tiling and then mapped to the triangular (HEX) lattice. At this step, OD matrix is generated and the network is then optimized. \textbf{B)} Resulting optimal network state with its distribution of edges weights showing a bi-modal shape. Only fast edges having weight larger than a threshold ($w_{e} > 5$) are kept to isolate the sub-graph constituted by a high velocity set of edges, such as a subway system.
    }}
    \label{fig:hex_mapping}
\end{figure}

\begin{figure}[H]
    \centering
    \includegraphics[width=0.9\linewidth]{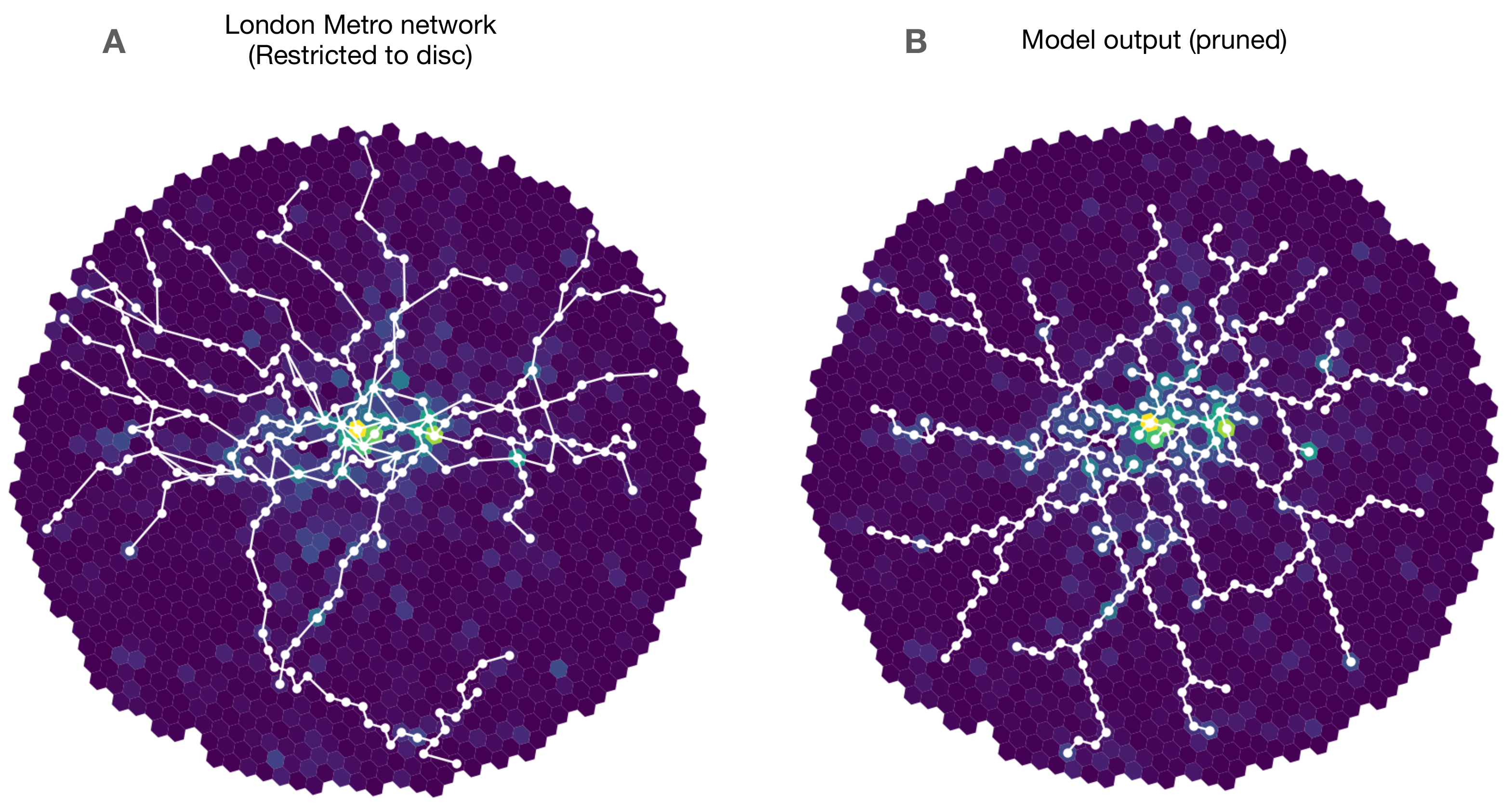}
    \caption{
    \small{
        \textbf{Pruned network vs real tube.}
    \textbf{A)} London subway network, restricted to the disc under study. \textbf{B)} GLA model output, limited to the nodes and edges which have a weight $w_{e} > 5.0$, where this value was chosen from the previously plotted $\{w_{e}\}$ distribution as a threshold to separate the second ``mode" with fast edges. This ``discretized" network is the model on which statistical measures are performed.
    }}
    \label{fig:pruned_network}
\end{figure}

\newpage

\section{Scaling of network stations}

The spatial organization of the transportation network can be inspected by taking into consideration the number of nodes' stations $N(r)$ up to a distance $r$ from the barycenter of stations. In the Main text we analyze the scaling regimes for the simulated and real London Tube network \cite{Gallotti2015} following the analysis employed in \cite{Roth2012}. Specifically, they obtain functional forms for the scaling properties of $N(r)$ for different distances from the barycenter. We report here the scalings obtained in \cite{Roth2012}:

\begin{equation}
    N(r) \sim \begin{cases}\rho_C \pi r^2 & \text { for } r<r_C \\ \rho_C \pi r_C^2+\mathcal{N}_B \int_{r_C}^r \frac{d r}{\Delta(r)} & \text { for } \quad r_C<r<r_{\max } \\ N & \text { for } r>r_{\max }\end{cases}
    \label{eq:scalings}
\end{equation}

Specifically, they show that in the large distance regime ($r > r_{C}$ and $r < r_{\max}$ ) the number of stations can be approximated by adding the integral curve $\mathcal{N}_B \int_{r_C}^r \frac{d r}{\Delta(r)}$ to a constant term. In \cite{Roth2012} it is also reported that the large distance behavior can be also, in general, approximated by a scaling law. 
Therefore we plot here the computation of the integral curve against rescaled values of $r$, and show that in that regime it can be approximated by a power law, and its exponent can be obtained via a linear fit. The $N(r)$ curve in the Main text was computed on the real London Tube restricted to the disc area as presented in SM Fig. \ref{fig:pruned_network}.
The associated exponent $\gamma = 1.25 \pm 0.02$ was used in the Main text to highlight the secondary scaling for $r > r_{C}$. As discussed in Ref. \cite{Roth2012}, due to $\Delta(r)$ in SM Eq. \ref{eq:scalings} being often noisy, this scaling property is not often well reproduced in empirical networks, and it is often restricted to a small region of $r$ values. In Fig. 6 of the Main text, we see that the secondary scaling with the exponent $\gamma$ approximates $N(r)$ in a limited interval for $r > r_{C}$. The value of the exponent in the branches region is expected to be $\gamma < 2.0$ \cite{Roth2012} and our result is consistent with this.

\begin{figure}[H]
    \centering
    \includegraphics[width=0.7\linewidth]{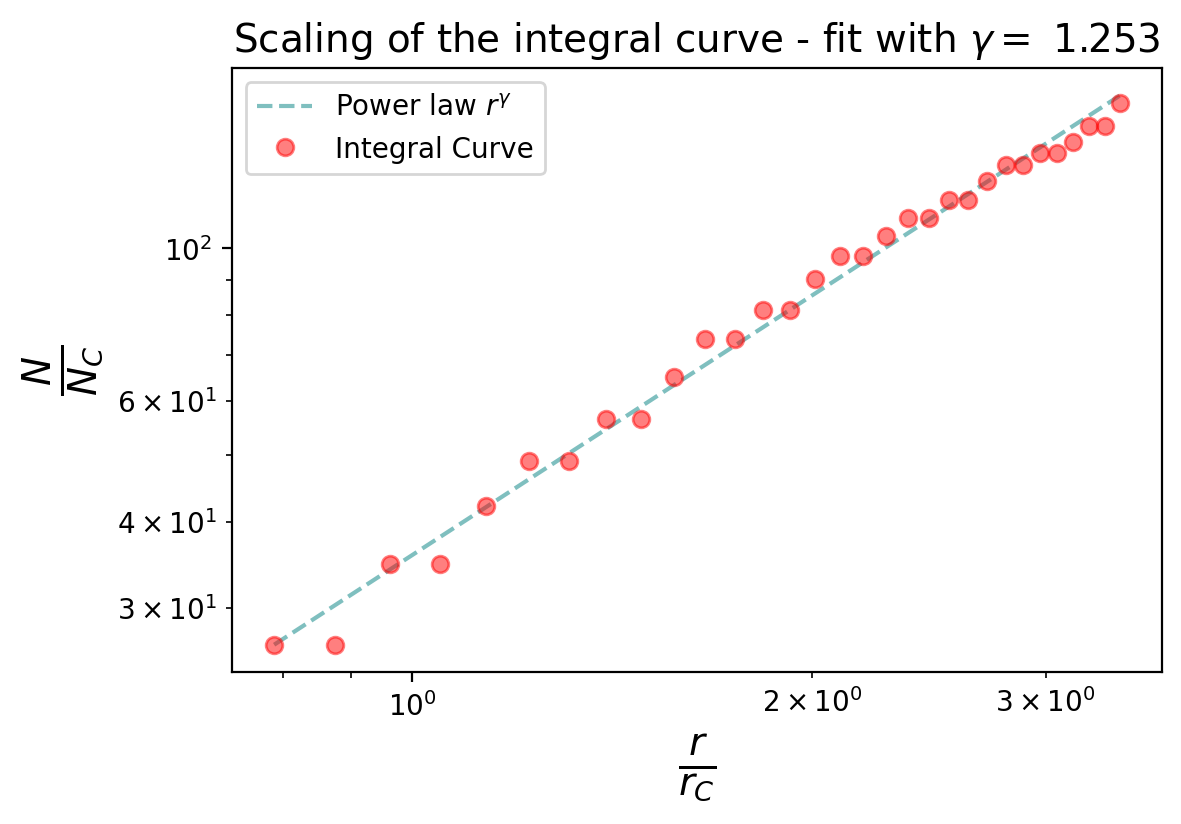}
    \caption{
    \small{
        \textbf{Scaling of the integral curve.}
        Log-Log plot of the integral curve and its linear fit. With the associated value of the $\gamma$ exponent approximated, to highlight a secondary scaling region as mentioned in \cite{Roth2012}.
    }}
    \label{fig:scaling-integral}
\end{figure}

\clearpage
\bibliography{apssamp}

\begin{thebibliography}{40}%
\makeatletter
\providecommand \@ifxundefined [1]{%
 \@ifx{#1\undefined}
}%
\providecommand \@ifnum [1]{%
 \ifnum #1\expandafter \@firstoftwo
 \else \expandafter \@secondoftwo
 \fi
}%
\providecommand \@ifx [1]{%
 \ifx #1\expandafter \@firstoftwo
 \else \expandafter \@secondoftwo
 \fi
}%
\providecommand \natexlab [1]{#1}%
\providecommand \enquote  [1]{``#1''}%
\providecommand \bibnamefont  [1]{#1}%
\providecommand \bibfnamefont [1]{#1}%
\providecommand \citenamefont [1]{#1}%
\providecommand \href@noop [0]{\@secondoftwo}%
\providecommand \href [0]{\begingroup \@sanitize@url \@href}%
\providecommand \@href[1]{\@@startlink{#1}\@@href}%
\providecommand \@@href[1]{\endgroup#1\@@endlink}%
\providecommand \@sanitize@url [0]{\catcode `\\12\catcode `\$12\catcode
  `\&12\catcode `\#12\catcode `\^12\catcode `\_12\catcode `\%12\relax}%
\providecommand \@@startlink[1]{}%
\providecommand \@@endlink[0]{}%
\providecommand \url  [0]{\begingroup\@sanitize@url \@url }%
\providecommand \@url [1]{\endgroup\@href {#1}{\urlprefix }}%
\providecommand \urlprefix  [0]{URL }%
\providecommand \Eprint [0]{\href }%
\providecommand \doibase [0]{http://dx.doi.org/}%
\providecommand \selectlanguage [0]{\@gobble}%
\providecommand \bibinfo  [0]{\@secondoftwo}%
\providecommand \bibfield  [0]{\@secondoftwo}%
\providecommand \translation [1]{[#1]}%
\providecommand \BibitemOpen [0]{}%
\providecommand \bibitemStop [0]{}%
\providecommand \bibitemNoStop [0]{.\EOS\space}%
\providecommand \EOS [0]{\spacefactor3000\relax}%
\providecommand \BibitemShut  [1]{\csname bibitem#1\endcsname}%
\let\auto@bib@innerbib\@empty
\bibitem [{\citenamefont {Batty}(2008)}]{Batty_2008}%
  \BibitemOpen
  \bibfield  {author} {\bibinfo {author} {\bibfnamefont {M.}~\bibnamefont
  {Batty}},\ }\href {\doibase 10.1126/science.1151419} {\bibfield  {journal}
  {\bibinfo  {journal} {Science}\ }\textbf {\bibinfo {volume} {319}},\ \bibinfo
  {pages} {769} (\bibinfo {year} {2008})}\BibitemShut {NoStop}%
\bibitem [{\citenamefont {Barthelemy}(2019)}]{Barthelemy_2019}%
  \BibitemOpen
  \bibfield  {author} {\bibinfo {author} {\bibfnamefont {M.}~\bibnamefont
  {Barthelemy}},\ }\href {\doibase 10.1038/s42254-019-0054-2} {\bibfield
  {journal} {\bibinfo  {journal} {Nature Reviews Physics}\ }\textbf {\bibinfo
  {volume} {1}},\ \bibinfo {pages} {406} (\bibinfo {year} {2019})}\BibitemShut
  {NoStop}%
\bibitem [{\citenamefont {Bettencourt}\ and\ \citenamefont
  {West}(2010)}]{Bettencourt2010}%
  \BibitemOpen
  \bibfield  {author} {\bibinfo {author} {\bibfnamefont {L.}~\bibnamefont
  {Bettencourt}}\ and\ \bibinfo {author} {\bibfnamefont {G.}~\bibnamefont
  {West}},\ }\href {\doibase 10.1038/467912a} {\bibfield  {journal} {\bibinfo
  {journal} {Nature}\ }\textbf {\bibinfo {volume} {467}},\ \bibinfo {pages}
  {912} (\bibinfo {year} {2010})}\BibitemShut {NoStop}%
\bibitem [{\citenamefont {Pan}\ \emph {et~al.}(2013)\citenamefont {Pan},
  \citenamefont {Ghoshal}, \citenamefont {Krumme}, \citenamefont {Cebrian},\
  and\ \citenamefont {Pentland}}]{Pan2013}%
  \BibitemOpen
  \bibfield  {author} {\bibinfo {author} {\bibfnamefont {W.}~\bibnamefont
  {Pan}}, \bibinfo {author} {\bibfnamefont {G.}~\bibnamefont {Ghoshal}},
  \bibinfo {author} {\bibfnamefont {C.}~\bibnamefont {Krumme}}, \bibinfo
  {author} {\bibfnamefont {M.}~\bibnamefont {Cebrian}}, \ and\ \bibinfo
  {author} {\bibfnamefont {A.}~\bibnamefont {Pentland}},\ }\href {\doibase
  10.1038/ncomms2961} {\bibfield  {journal} {\bibinfo  {journal} {Nature
  Communications}\ }\textbf {\bibinfo {volume} {4}} (\bibinfo {year} {2013}),\
  10.1038/ncomms2961}\BibitemShut {NoStop}%
\bibitem [{\citenamefont {Arcaute}\ \emph {et~al.}(2015)\citenamefont
  {Arcaute}, \citenamefont {Hatna}, \citenamefont {Ferguson}, \citenamefont
  {Youn}, \citenamefont {Johansson},\ and\ \citenamefont
  {Batty}}]{Arcaute2015}%
  \BibitemOpen
  \bibfield  {author} {\bibinfo {author} {\bibfnamefont {E.}~\bibnamefont
  {Arcaute}}, \bibinfo {author} {\bibfnamefont {E.}~\bibnamefont {Hatna}},
  \bibinfo {author} {\bibfnamefont {P.}~\bibnamefont {Ferguson}}, \bibinfo
  {author} {\bibfnamefont {H.}~\bibnamefont {Youn}}, \bibinfo {author}
  {\bibfnamefont {A.}~\bibnamefont {Johansson}}, \ and\ \bibinfo {author}
  {\bibfnamefont {M.}~\bibnamefont {Batty}},\ }\href {\doibase
  10.1098/rsif.2014.0745} {\bibfield  {journal} {\bibinfo  {journal} {Journal
  of The Royal Society Interface}\ }\textbf {\bibinfo {volume} {12}},\ \bibinfo
  {pages} {20140745} (\bibinfo {year} {2015})}\BibitemShut {NoStop}%
\bibitem [{\citenamefont {Bettencourt}(2013)}]{Bettencourt2013}%
  \BibitemOpen
  \bibfield  {author} {\bibinfo {author} {\bibfnamefont {L.~M.~A.}\
  \bibnamefont {Bettencourt}},\ }\href {\doibase 10.1126/science.1235823}
  {\bibfield  {journal} {\bibinfo  {journal} {Science}\ }\textbf {\bibinfo
  {volume} {340}},\ \bibinfo {pages} {1438} (\bibinfo {year}
  {2013})}\BibitemShut {NoStop}%
\bibitem [{\citenamefont {Bassolas}\ \emph {et~al.}(2019)\citenamefont
  {Bassolas}, \citenamefont {Barbosa-Filho}, \citenamefont {Dickinson},
  \citenamefont {Dotiwalla}, \citenamefont {Eastham}, \citenamefont {Gallotti},
  \citenamefont {Ghoshal}, \citenamefont {Gipson}, \citenamefont {Hazarie},
  \citenamefont {Kautz}, \citenamefont {Kucuktunc}, \citenamefont {Lieber},
  \citenamefont {Sadilek},\ and\ \citenamefont {Ramasco}}]{Bassolas2019}%
  \BibitemOpen
  \bibfield  {author} {\bibinfo {author} {\bibfnamefont {A.}~\bibnamefont
  {Bassolas}}, \bibinfo {author} {\bibfnamefont {H.}~\bibnamefont
  {Barbosa-Filho}}, \bibinfo {author} {\bibfnamefont {B.}~\bibnamefont
  {Dickinson}}, \bibinfo {author} {\bibfnamefont {X.}~\bibnamefont
  {Dotiwalla}}, \bibinfo {author} {\bibfnamefont {P.}~\bibnamefont {Eastham}},
  \bibinfo {author} {\bibfnamefont {R.}~\bibnamefont {Gallotti}}, \bibinfo
  {author} {\bibfnamefont {G.}~\bibnamefont {Ghoshal}}, \bibinfo {author}
  {\bibfnamefont {B.}~\bibnamefont {Gipson}}, \bibinfo {author} {\bibfnamefont
  {S.~A.}\ \bibnamefont {Hazarie}}, \bibinfo {author} {\bibfnamefont
  {H.}~\bibnamefont {Kautz}}, \bibinfo {author} {\bibfnamefont
  {O.}~\bibnamefont {Kucuktunc}}, \bibinfo {author} {\bibfnamefont
  {A.}~\bibnamefont {Lieber}}, \bibinfo {author} {\bibfnamefont
  {A.}~\bibnamefont {Sadilek}}, \ and\ \bibinfo {author} {\bibfnamefont
  {J.~J.}\ \bibnamefont {Ramasco}},\ }\href {\doibase
  10.1038/s41467-019-12809-y} {\bibfield  {journal} {\bibinfo  {journal}
  {Nature Communications}\ }\textbf {\bibinfo {volume} {10}} (\bibinfo {year}
  {2019}),\ 10.1038/s41467-019-12809-y}\BibitemShut {NoStop}%
\bibitem [{\citenamefont {Bettencourt}(2020)}]{Bettencourt2020}%
  \BibitemOpen
  \bibfield  {author} {\bibinfo {author} {\bibfnamefont {L.~M.~A.}\
  \bibnamefont {Bettencourt}},\ }\href {\doibase 10.1126/sciadv.aat8812}
  {\bibfield  {journal} {\bibinfo  {journal} {Science Advances}\ }\textbf
  {\bibinfo {volume} {6}} (\bibinfo {year} {2020}),\
  10.1126/sciadv.aat8812}\BibitemShut {NoStop}%
\bibitem [{\citenamefont {Schl\"{a}pfer}\ \emph {et~al.}(2021)\citenamefont
  {Schl\"{a}pfer}, \citenamefont {Dong}, \citenamefont {O'Keeffe},
  \citenamefont {Santi}, \citenamefont {Szell}, \citenamefont {Salat},
  \citenamefont {Anklesaria}, \citenamefont {Vazifeh}, \citenamefont {Ratti},\
  and\ \citenamefont {West}}]{Schlpfer2021}%
  \BibitemOpen
  \bibfield  {author} {\bibinfo {author} {\bibfnamefont {M.}~\bibnamefont
  {Schl\"{a}pfer}}, \bibinfo {author} {\bibfnamefont {L.}~\bibnamefont {Dong}},
  \bibinfo {author} {\bibfnamefont {K.}~\bibnamefont {O'Keeffe}}, \bibinfo
  {author} {\bibfnamefont {P.}~\bibnamefont {Santi}}, \bibinfo {author}
  {\bibfnamefont {M.}~\bibnamefont {Szell}}, \bibinfo {author} {\bibfnamefont
  {H.}~\bibnamefont {Salat}}, \bibinfo {author} {\bibfnamefont
  {S.}~\bibnamefont {Anklesaria}}, \bibinfo {author} {\bibfnamefont
  {M.}~\bibnamefont {Vazifeh}}, \bibinfo {author} {\bibfnamefont
  {C.}~\bibnamefont {Ratti}}, \ and\ \bibinfo {author} {\bibfnamefont {G.~B.}\
  \bibnamefont {West}},\ }\href {\doibase 10.1038/s41586-021-03480-9}
  {\bibfield  {journal} {\bibinfo  {journal} {Nature}\ }\textbf {\bibinfo
  {volume} {593}},\ \bibinfo {pages} {522} (\bibinfo {year}
  {2021})}\BibitemShut {NoStop}%
\bibitem [{\citenamefont {Barbosa}\ \emph {et~al.}(2018)\citenamefont
  {Barbosa}, \citenamefont {Barthelemy}, \citenamefont {Ghoshal}, \citenamefont
  {James}, \citenamefont {Lenormand}, \citenamefont {Louail}, \citenamefont
  {Menezes}, \citenamefont {Ramasco}, \citenamefont {Simini},\ and\
  \citenamefont {Tomasini}}]{Barbosa2018}%
  \BibitemOpen
  \bibfield  {author} {\bibinfo {author} {\bibfnamefont {H.}~\bibnamefont
  {Barbosa}}, \bibinfo {author} {\bibfnamefont {M.}~\bibnamefont {Barthelemy}},
  \bibinfo {author} {\bibfnamefont {G.}~\bibnamefont {Ghoshal}}, \bibinfo
  {author} {\bibfnamefont {C.~R.}\ \bibnamefont {James}}, \bibinfo {author}
  {\bibfnamefont {M.}~\bibnamefont {Lenormand}}, \bibinfo {author}
  {\bibfnamefont {T.}~\bibnamefont {Louail}}, \bibinfo {author} {\bibfnamefont
  {R.}~\bibnamefont {Menezes}}, \bibinfo {author} {\bibfnamefont {J.~J.}\
  \bibnamefont {Ramasco}}, \bibinfo {author} {\bibfnamefont {F.}~\bibnamefont
  {Simini}}, \ and\ \bibinfo {author} {\bibfnamefont {M.}~\bibnamefont
  {Tomasini}},\ }\href {\doibase 10.1016/j.physrep.2018.01.001} {\bibfield
  {journal} {\bibinfo  {journal} {Physics Reports}\ }\textbf {\bibinfo {volume}
  {734}},\ \bibinfo {pages} {1} (\bibinfo {year} {2018})}\BibitemShut {NoStop}%
\bibitem [{\citenamefont {Alessandretti}\ \emph {et~al.}(2022)\citenamefont
  {Alessandretti}, \citenamefont {Orozco}, \citenamefont {Saberi},
  \citenamefont {Szell},\ and\ \citenamefont {Battiston}}]{Alessandretti2022}%
  \BibitemOpen
  \bibfield  {author} {\bibinfo {author} {\bibfnamefont {L.}~\bibnamefont
  {Alessandretti}}, \bibinfo {author} {\bibfnamefont {L.~G.~N.}\ \bibnamefont
  {Orozco}}, \bibinfo {author} {\bibfnamefont {M.}~\bibnamefont {Saberi}},
  \bibinfo {author} {\bibfnamefont {M.}~\bibnamefont {Szell}}, \ and\ \bibinfo
  {author} {\bibfnamefont {F.}~\bibnamefont {Battiston}},\ }\href {\doibase
  10.1177/23998083221108190} {\bibfield  {journal} {\bibinfo  {journal}
  {Environment and Planning B: Urban Analytics and City Science}\ ,\ \bibinfo
  {pages} {239980832211081}} (\bibinfo {year} {2022})}\BibitemShut {NoStop}%
\bibitem [{\citenamefont {Lee}\ \emph {et~al.}(2017)\citenamefont {Lee},
  \citenamefont {Barbosa}, \citenamefont {Youn}, \citenamefont {Holme},\ and\
  \citenamefont {Ghoshal}}]{Lee2017}%
  \BibitemOpen
  \bibfield  {author} {\bibinfo {author} {\bibfnamefont {M.}~\bibnamefont
  {Lee}}, \bibinfo {author} {\bibfnamefont {H.}~\bibnamefont {Barbosa}},
  \bibinfo {author} {\bibfnamefont {H.}~\bibnamefont {Youn}}, \bibinfo {author}
  {\bibfnamefont {P.}~\bibnamefont {Holme}}, \ and\ \bibinfo {author}
  {\bibfnamefont {G.}~\bibnamefont {Ghoshal}},\ }\href {\doibase
  10.1038/s41467-017-02374-7} {\bibfield  {journal} {\bibinfo  {journal}
  {Nature Communications}\ }\textbf {\bibinfo {volume} {8}} (\bibinfo {year}
  {2017}),\ 10.1038/s41467-017-02374-7}\BibitemShut {NoStop}%
\bibitem [{\citenamefont {Gallotti}\ \emph {et~al.}(2021)\citenamefont
  {Gallotti}, \citenamefont {Sacco},\ and\ \citenamefont
  {Domenico}}]{Gallotti2021}%
  \BibitemOpen
  \bibfield  {author} {\bibinfo {author} {\bibfnamefont {R.}~\bibnamefont
  {Gallotti}}, \bibinfo {author} {\bibfnamefont {P.}~\bibnamefont {Sacco}}, \
  and\ \bibinfo {author} {\bibfnamefont {M.~D.}\ \bibnamefont {Domenico}},\
  }\href {\doibase 10.1155/2021/1782354} {\bibfield  {journal} {\bibinfo
  {journal} {Complexity}\ }\textbf {\bibinfo {volume} {2021}},\ \bibinfo
  {pages} {1} (\bibinfo {year} {2021})}\BibitemShut {NoStop}%
\bibitem [{\citenamefont {Barth{\'{e}}lemy}(2011)}]{Barthelemy_2011}%
  \BibitemOpen
  \bibfield  {author} {\bibinfo {author} {\bibfnamefont {M.}~\bibnamefont
  {Barth{\'{e}}lemy}},\ }\href {\doibase 10.1016/j.physrep.2010.11.002}
  {\bibfield  {journal} {\bibinfo  {journal} {Physics Reports}\ }\textbf
  {\bibinfo {volume} {499}},\ \bibinfo {pages} {1} (\bibinfo {year}
  {2011})}\BibitemShut {NoStop}%
\bibitem [{\citenamefont {Morris}\ and\ \citenamefont
  {Barthelemy}(2012)}]{Morris2012}%
  \BibitemOpen
  \bibfield  {author} {\bibinfo {author} {\bibfnamefont {R.~G.}\ \bibnamefont
  {Morris}}\ and\ \bibinfo {author} {\bibfnamefont {M.}~\bibnamefont
  {Barthelemy}},\ }\href {\doibase 10.1103/physrevlett.109.128703} {\bibfield
  {journal} {\bibinfo  {journal} {Physical Review Letters}\ }\textbf {\bibinfo
  {volume} {109}} (\bibinfo {year} {2012}),\
  10.1103/physrevlett.109.128703}\BibitemShut {NoStop}%
\bibitem [{\citenamefont {Gastner}\ and\ \citenamefont
  {Newman}(2006)}]{Gastner_2006}%
  \BibitemOpen
  \bibfield  {author} {\bibinfo {author} {\bibfnamefont {M.~T.}\ \bibnamefont
  {Gastner}}\ and\ \bibinfo {author} {\bibfnamefont {M.~E.~J.}\ \bibnamefont
  {Newman}},\ }\href {\doibase 10.1103/physreve.74.016117} {\bibfield
  {journal} {\bibinfo  {journal} {Physical Review E}\ }\textbf {\bibinfo
  {volume} {74}} (\bibinfo {year} {2006}),\
  10.1103/physreve.74.016117}\BibitemShut {NoStop}%
\bibitem [{\citenamefont {Barth{\'{e}}lemy}\ and\ \citenamefont
  {Flammini}(2006)}]{Barthelemy2006}%
  \BibitemOpen
  \bibfield  {author} {\bibinfo {author} {\bibfnamefont {M.}~\bibnamefont
  {Barth{\'{e}}lemy}}\ and\ \bibinfo {author} {\bibfnamefont {A.}~\bibnamefont
  {Flammini}},\ }\href {\doibase 10.1088/1742-5468/2006/07/l07002} {\bibfield
  {journal} {\bibinfo  {journal} {Journal of Statistical Mechanics: Theory and
  Experiment}\ }\textbf {\bibinfo {volume} {2006}},\ \bibinfo {pages} {L07002}
  (\bibinfo {year} {2006})}\BibitemShut {NoStop}%
\bibitem [{\citenamefont {Louf}\ \emph {et~al.}(2013)\citenamefont {Louf},
  \citenamefont {Jensen},\ and\ \citenamefont {Barthelemy}}]{Louf_2013}%
  \BibitemOpen
  \bibfield  {author} {\bibinfo {author} {\bibfnamefont {R.}~\bibnamefont
  {Louf}}, \bibinfo {author} {\bibfnamefont {P.}~\bibnamefont {Jensen}}, \ and\
  \bibinfo {author} {\bibfnamefont {M.}~\bibnamefont {Barthelemy}},\ }\href
  {\doibase 10.1073/pnas.1222441110} {\bibfield  {journal} {\bibinfo  {journal}
  {Proceedings of the National Academy of Sciences}\ }\textbf {\bibinfo
  {volume} {110}},\ \bibinfo {pages} {8824} (\bibinfo {year}
  {2013})}\BibitemShut {NoStop}%
\bibitem [{\citenamefont {Banavar}\ \emph {et~al.}(1999)\citenamefont
  {Banavar}, \citenamefont {Maritan},\ and\ \citenamefont
  {Rinaldo}}]{Banavar1999}%
  \BibitemOpen
  \bibfield  {author} {\bibinfo {author} {\bibfnamefont {J.~R.}\ \bibnamefont
  {Banavar}}, \bibinfo {author} {\bibfnamefont {A.}~\bibnamefont {Maritan}}, \
  and\ \bibinfo {author} {\bibfnamefont {A.}~\bibnamefont {Rinaldo}},\ }\href
  {\doibase 10.1038/20144} {\bibfield  {journal} {\bibinfo  {journal} {Nature}\
  }\textbf {\bibinfo {volume} {399}},\ \bibinfo {pages} {130} (\bibinfo {year}
  {1999})}\BibitemShut {NoStop}%
\bibitem [{\citenamefont {Pei}\ \emph {et~al.}(2022)\citenamefont {Pei},
  \citenamefont {Xiao}, \citenamefont {Yu},\ and\ \citenamefont
  {Li}}]{Pei_2022}%
  \BibitemOpen
  \bibfield  {author} {\bibinfo {author} {\bibfnamefont {A.}~\bibnamefont
  {Pei}}, \bibinfo {author} {\bibfnamefont {F.}~\bibnamefont {Xiao}}, \bibinfo
  {author} {\bibfnamefont {S.}~\bibnamefont {Yu}}, \ and\ \bibinfo {author}
  {\bibfnamefont {L.}~\bibnamefont {Li}},\ }\href {\doibase
  10.1038/s41598-022-12053-3} {\bibfield  {journal} {\bibinfo  {journal}
  {Scientific Reports}\ }\textbf {\bibinfo {volume} {12}} (\bibinfo {year}
  {2022}),\ 10.1038/s41598-022-12053-3}\BibitemShut {NoStop}%
\bibitem [{\citenamefont {Gallotti}\ and\ \citenamefont
  {Barthelemy}(2015)}]{Gallotti2015}%
  \BibitemOpen
  \bibfield  {author} {\bibinfo {author} {\bibfnamefont {R.}~\bibnamefont
  {Gallotti}}\ and\ \bibinfo {author} {\bibfnamefont {M.}~\bibnamefont
  {Barthelemy}},\ }\href {\doibase 10.1038/sdata.2014.56} {\bibfield  {journal}
  {\bibinfo  {journal} {Scientific Data}\ }\textbf {\bibinfo {volume} {2}}
  (\bibinfo {year} {2015}),\ 10.1038/sdata.2014.56}\BibitemShut {NoStop}%
\bibitem [{\citenamefont {Roth}\ \emph {et~al.}(2012)\citenamefont {Roth},
  \citenamefont {Kang}, \citenamefont {Batty},\ and\ \citenamefont
  {Barthelemy}}]{Roth2012}%
  \BibitemOpen
  \bibfield  {author} {\bibinfo {author} {\bibfnamefont {C.}~\bibnamefont
  {Roth}}, \bibinfo {author} {\bibfnamefont {S.~M.}\ \bibnamefont {Kang}},
  \bibinfo {author} {\bibfnamefont {M.}~\bibnamefont {Batty}}, \ and\ \bibinfo
  {author} {\bibfnamefont {M.}~\bibnamefont {Barthelemy}},\ }\href {\doibase
  10.1098/rsif.2012.0259} {\bibfield  {journal} {\bibinfo  {journal} {Journal
  of The Royal Society Interface}\ }\textbf {\bibinfo {volume} {9}},\ \bibinfo
  {pages} {2540} (\bibinfo {year} {2012})}\BibitemShut {NoStop}%
\bibitem [{\citenamefont {Tero}\ \emph {et~al.}(2010)\citenamefont {Tero},
  \citenamefont {Takagi}, \citenamefont {Saigusa}, \citenamefont {Ito},
  \citenamefont {Bebber}, \citenamefont {Fricker}, \citenamefont {Yumiki},
  \citenamefont {Kobayashi},\ and\ \citenamefont {Nakagaki}}]{Tero2010}%
  \BibitemOpen
  \bibfield  {author} {\bibinfo {author} {\bibfnamefont {A.}~\bibnamefont
  {Tero}}, \bibinfo {author} {\bibfnamefont {S.}~\bibnamefont {Takagi}},
  \bibinfo {author} {\bibfnamefont {T.}~\bibnamefont {Saigusa}}, \bibinfo
  {author} {\bibfnamefont {K.}~\bibnamefont {Ito}}, \bibinfo {author}
  {\bibfnamefont {D.~P.}\ \bibnamefont {Bebber}}, \bibinfo {author}
  {\bibfnamefont {M.~D.}\ \bibnamefont {Fricker}}, \bibinfo {author}
  {\bibfnamefont {K.}~\bibnamefont {Yumiki}}, \bibinfo {author} {\bibfnamefont
  {R.}~\bibnamefont {Kobayashi}}, \ and\ \bibinfo {author} {\bibfnamefont
  {T.}~\bibnamefont {Nakagaki}},\ }\href {\doibase 10.1126/science.1177894}
  {\bibfield  {journal} {\bibinfo  {journal} {Science}\ }\textbf {\bibinfo
  {volume} {327}},\ \bibinfo {pages} {439} (\bibinfo {year}
  {2010})}\BibitemShut {NoStop}%
\bibitem [{\citenamefont {Louf}\ and\ \citenamefont
  {Barthelemy}(2014)}]{Louf_2014}%
  \BibitemOpen
  \bibfield  {author} {\bibinfo {author} {\bibfnamefont {R.}~\bibnamefont
  {Louf}}\ and\ \bibinfo {author} {\bibfnamefont {M.}~\bibnamefont
  {Barthelemy}},\ }\href {\doibase 10.1038/srep05561} {\bibfield  {journal}
  {\bibinfo  {journal} {Scientific Reports}\ }\textbf {\bibinfo {volume} {4}}
  (\bibinfo {year} {2014}),\ 10.1038/srep05561}\BibitemShut {NoStop}%
\bibitem [{\citenamefont {Ibrahim}\ \emph {et~al.}(2021)\citenamefont
  {Ibrahim}, \citenamefont {Lonardi},\ and\ \citenamefont
  {Bacco}}]{Ibrahim2021}%
  \BibitemOpen
  \bibfield  {author} {\bibinfo {author} {\bibfnamefont {A.~A.}\ \bibnamefont
  {Ibrahim}}, \bibinfo {author} {\bibfnamefont {A.}~\bibnamefont {Lonardi}}, \
  and\ \bibinfo {author} {\bibfnamefont {C.~D.}\ \bibnamefont {Bacco}},\ }\href
  {\doibase 10.3390/a14070189} {\bibfield  {journal} {\bibinfo  {journal}
  {Algorithms}\ }\textbf {\bibinfo {volume} {14}},\ \bibinfo {pages} {189}
  (\bibinfo {year} {2021})}\BibitemShut {NoStop}%
\bibitem [{\citenamefont {Zhang}\ \emph {et~al.}(2015)\citenamefont {Zhang},
  \citenamefont {Adamatzky}, \citenamefont {Chan}, \citenamefont {Deng},
  \citenamefont {Yang}, \citenamefont {Yang}, \citenamefont {Tsompanas},
  \citenamefont {Sirakoulis},\ and\ \citenamefont {Mahadevan}}]{Zhang2015}%
  \BibitemOpen
  \bibfield  {author} {\bibinfo {author} {\bibfnamefont {X.}~\bibnamefont
  {Zhang}}, \bibinfo {author} {\bibfnamefont {A.}~\bibnamefont {Adamatzky}},
  \bibinfo {author} {\bibfnamefont {F.~T.}\ \bibnamefont {Chan}}, \bibinfo
  {author} {\bibfnamefont {Y.}~\bibnamefont {Deng}}, \bibinfo {author}
  {\bibfnamefont {H.}~\bibnamefont {Yang}}, \bibinfo {author} {\bibfnamefont
  {X.-S.}\ \bibnamefont {Yang}}, \bibinfo {author} {\bibfnamefont {M.-A.~I.}\
  \bibnamefont {Tsompanas}}, \bibinfo {author} {\bibfnamefont {G.~C.}\
  \bibnamefont {Sirakoulis}}, \ and\ \bibinfo {author} {\bibfnamefont
  {S.}~\bibnamefont {Mahadevan}},\ }\href {\doibase 10.1038/srep10794}
  {\bibfield  {journal} {\bibinfo  {journal} {Scientific Reports}\ }\textbf
  {\bibinfo {volume} {5}} (\bibinfo {year} {2015}),\
  10.1038/srep10794}\BibitemShut {NoStop}%
\bibitem [{\citenamefont {Szell}\ \emph {et~al.}(2022)\citenamefont {Szell},
  \citenamefont {Mimar}, \citenamefont {Perlman}, \citenamefont {Ghoshal},\
  and\ \citenamefont {Sinatra}}]{Szell2022}%
  \BibitemOpen
  \bibfield  {author} {\bibinfo {author} {\bibfnamefont {M.}~\bibnamefont
  {Szell}}, \bibinfo {author} {\bibfnamefont {S.}~\bibnamefont {Mimar}},
  \bibinfo {author} {\bibfnamefont {T.}~\bibnamefont {Perlman}}, \bibinfo
  {author} {\bibfnamefont {G.}~\bibnamefont {Ghoshal}}, \ and\ \bibinfo
  {author} {\bibfnamefont {R.}~\bibnamefont {Sinatra}},\ }\href {\doibase
  10.1038/s41598-022-10783-y} {\bibfield  {journal} {\bibinfo  {journal}
  {Scientific Reports}\ }\textbf {\bibinfo {volume} {12}} (\bibinfo {year}
  {2022}),\ 10.1038/s41598-022-10783-y}\BibitemShut {NoStop}%
\bibitem [{\citenamefont {Birch}\ \emph {et~al.}(2007)\citenamefont {Birch},
  \citenamefont {Oom},\ and\ \citenamefont {Beecham}}]{Birch2007}%
  \BibitemOpen
  \bibfield  {author} {\bibinfo {author} {\bibfnamefont {C.~P.}\ \bibnamefont
  {Birch}}, \bibinfo {author} {\bibfnamefont {S.~P.}\ \bibnamefont {Oom}}, \
  and\ \bibinfo {author} {\bibfnamefont {J.~A.}\ \bibnamefont {Beecham}},\
  }\href {\doibase 10.1016/j.ecolmodel.2007.03.041} {\bibfield  {journal}
  {\bibinfo  {journal} {Ecological Modelling}\ }\textbf {\bibinfo {volume}
  {206}},\ \bibinfo {pages} {347} (\bibinfo {year} {2007})}\BibitemShut
  {NoStop}%
\bibitem [{\citenamefont {Gallotti}\ \emph {et~al.}(2016)\citenamefont
  {Gallotti}, \citenamefont {Bazzani}, \citenamefont {Rambaldi},\ and\
  \citenamefont {Barthelemy}}]{Gallotti2016}%
  \BibitemOpen
  \bibfield  {author} {\bibinfo {author} {\bibfnamefont {R.}~\bibnamefont
  {Gallotti}}, \bibinfo {author} {\bibfnamefont {A.}~\bibnamefont {Bazzani}},
  \bibinfo {author} {\bibfnamefont {S.}~\bibnamefont {Rambaldi}}, \ and\
  \bibinfo {author} {\bibfnamefont {M.}~\bibnamefont {Barthelemy}},\ }\href
  {\doibase 10.1038/ncomms12600} {\bibfield  {journal} {\bibinfo  {journal}
  {Nature Communications}\ }\textbf {\bibinfo {volume} {7}} (\bibinfo {year}
  {2016}),\ 10.1038/ncomms12600}\BibitemShut {NoStop}%
\bibitem [{\citenamefont {Ren}\ \emph {et~al.}(2014)\citenamefont {Ren},
  \citenamefont {Ercsey-Ravasz}, \citenamefont {Wang}, \citenamefont
  {Gonz{\'{a}}lez},\ and\ \citenamefont {Toroczkai}}]{Ren_2014}%
  \BibitemOpen
  \bibfield  {author} {\bibinfo {author} {\bibfnamefont {Y.}~\bibnamefont
  {Ren}}, \bibinfo {author} {\bibfnamefont {M.}~\bibnamefont {Ercsey-Ravasz}},
  \bibinfo {author} {\bibfnamefont {P.}~\bibnamefont {Wang}}, \bibinfo {author}
  {\bibfnamefont {M.~C.}\ \bibnamefont {Gonz{\'{a}}lez}}, \ and\ \bibinfo
  {author} {\bibfnamefont {Z.}~\bibnamefont {Toroczkai}},\ }\href {\doibase
  10.1038/ncomms6347} {\bibfield  {journal} {\bibinfo  {journal} {Nature
  Communications}\ }\textbf {\bibinfo {volume} {5}} (\bibinfo {year} {2014}),\
  10.1038/ncomms6347}\BibitemShut {NoStop}%
\bibitem [{\citenamefont {Wilson}(1975)}]{Wilson_1975}%
  \BibitemOpen
  \bibfield  {author} {\bibinfo {author} {\bibfnamefont {A.}~\bibnamefont
  {Wilson}},\ }\href {\doibase 10.1016/0041-1647(75)90054-4} {\bibfield
  {journal} {\bibinfo  {journal} {Transportation Research}\ }\textbf {\bibinfo
  {volume} {9}},\ \bibinfo {pages} {167} (\bibinfo {year} {1975})}\BibitemShut
  {NoStop}%
\bibitem [{\citenamefont {Viana}\ \emph {et~al.}(2013)\citenamefont {Viana},
  \citenamefont {Strano}, \citenamefont {Bordin},\ and\ \citenamefont
  {Barthelemy}}]{Viana2013}%
  \BibitemOpen
  \bibfield  {author} {\bibinfo {author} {\bibfnamefont {M.~P.}\ \bibnamefont
  {Viana}}, \bibinfo {author} {\bibfnamefont {E.}~\bibnamefont {Strano}},
  \bibinfo {author} {\bibfnamefont {P.}~\bibnamefont {Bordin}}, \ and\ \bibinfo
  {author} {\bibfnamefont {M.}~\bibnamefont {Barthelemy}},\ }\href {\doibase
  10.1038/srep03495} {\bibfield  {journal} {\bibinfo  {journal} {Scientific
  Reports}\ }\textbf {\bibinfo {volume} {3}} (\bibinfo {year} {2013}),\
  10.1038/srep03495}\BibitemShut {NoStop}%
\bibitem [{\citenamefont {Newman}(2010)}]{Newman2010}%
  \BibitemOpen
  \bibfield  {author} {\bibinfo {author} {\bibfnamefont {M.}~\bibnamefont
  {Newman}},\ }\href {\doibase 10.1093/acprof:oso/9780199206650.001.0001}
  {\emph {\bibinfo {title} {Networks}}}\ (\bibinfo  {publisher} {Oxford
  University Press},\ \bibinfo {year} {2010})\BibitemShut {NoStop}%
\bibitem [{\citenamefont {Dreyfus}\ and\ \citenamefont
  {Wagner}(1971)}]{Dreyfus1971}%
  \BibitemOpen
  \bibfield  {author} {\bibinfo {author} {\bibfnamefont {S.~E.}\ \bibnamefont
  {Dreyfus}}\ and\ \bibinfo {author} {\bibfnamefont {R.~A.}\ \bibnamefont
  {Wagner}},\ }\href {\doibase 10.1002/net.3230010302} {\bibfield  {journal}
  {\bibinfo  {journal} {Networks}\ }\textbf {\bibinfo {volume} {1}},\ \bibinfo
  {pages} {195} (\bibinfo {year} {1971})}\BibitemShut {NoStop}%
\bibitem [{\citenamefont {Brazil}\ \emph {et~al.}(2013)\citenamefont {Brazil},
  \citenamefont {Graham}, \citenamefont {Thomas},\ and\ \citenamefont
  {Zachariasen}}]{Brazil2013}%
  \BibitemOpen
  \bibfield  {author} {\bibinfo {author} {\bibfnamefont {M.}~\bibnamefont
  {Brazil}}, \bibinfo {author} {\bibfnamefont {R.~L.}\ \bibnamefont {Graham}},
  \bibinfo {author} {\bibfnamefont {D.~A.}\ \bibnamefont {Thomas}}, \ and\
  \bibinfo {author} {\bibfnamefont {M.}~\bibnamefont {Zachariasen}},\ }\href
  {\doibase 10.1007/s00407-013-0127-z} {\bibfield  {journal} {\bibinfo
  {journal} {Archive for History of Exact Sciences}\ }\textbf {\bibinfo
  {volume} {68}},\ \bibinfo {pages} {327} (\bibinfo {year} {2013})}\BibitemShut
  {NoStop}%
\bibitem [{\citenamefont {Kavitha}\ \emph {et~al.}(2007)\citenamefont
  {Kavitha}, \citenamefont {Mehlhorn}, \citenamefont {Michail},\ and\
  \citenamefont {Paluch}}]{Kavitha2007}%
  \BibitemOpen
  \bibfield  {author} {\bibinfo {author} {\bibfnamefont {T.}~\bibnamefont
  {Kavitha}}, \bibinfo {author} {\bibfnamefont {K.}~\bibnamefont {Mehlhorn}},
  \bibinfo {author} {\bibfnamefont {D.}~\bibnamefont {Michail}}, \ and\
  \bibinfo {author} {\bibfnamefont {K.~E.}\ \bibnamefont {Paluch}},\ }\href
  {\doibase 10.1007/s00453-007-9064-z} {\bibfield  {journal} {\bibinfo
  {journal} {Algorithmica}\ }\textbf {\bibinfo {volume} {52}},\ \bibinfo
  {pages} {333} (\bibinfo {year} {2007})}\BibitemShut {NoStop}%
\bibitem [{\citenamefont {Katifori}\ \emph {et~al.}(2010)\citenamefont
  {Katifori}, \citenamefont {Sz\"{o}ll{\H{o}}si},\ and\ \citenamefont
  {Magnasco}}]{Katifori2010}%
  \BibitemOpen
  \bibfield  {author} {\bibinfo {author} {\bibfnamefont {E.}~\bibnamefont
  {Katifori}}, \bibinfo {author} {\bibfnamefont {G.~J.}\ \bibnamefont
  {Sz\"{o}ll{\H{o}}si}}, \ and\ \bibinfo {author} {\bibfnamefont {M.~O.}\
  \bibnamefont {Magnasco}},\ }\href {\doibase 10.1103/physrevlett.104.048704}
  {\bibfield  {journal} {\bibinfo  {journal} {Physical Review Letters}\
  }\textbf {\bibinfo {volume} {104}} (\bibinfo {year} {2010}),\
  10.1103/physrevlett.104.048704}\BibitemShut {NoStop}%
\bibitem [{\citenamefont {Piovani}\ \emph {et~al.}(2017)\citenamefont
  {Piovani}, \citenamefont {Molinero},\ and\ \citenamefont
  {Wilson}}]{Piovani2017}%
  \BibitemOpen
  \bibfield  {author} {\bibinfo {author} {\bibfnamefont {D.}~\bibnamefont
  {Piovani}}, \bibinfo {author} {\bibfnamefont {C.}~\bibnamefont {Molinero}}, \
  and\ \bibinfo {author} {\bibfnamefont {A.}~\bibnamefont {Wilson}},\ }\href
  {\doibase 10.1371/journal.pone.0185787} {\bibfield  {journal} {\bibinfo
  {journal} {{PLOS} {ONE}}\ }\textbf {\bibinfo {volume} {12}},\ \bibinfo
  {pages} {e0185787} (\bibinfo {year} {2017})}\BibitemShut {NoStop}%
\bibitem [{\citenamefont {Hidalgo}\ \emph {et~al.}(2020)\citenamefont
  {Hidalgo}, \citenamefont {Casta{\~{n}}er},\ and\ \citenamefont
  {Sevtsuk}}]{Hidalgo_2020}%
  \BibitemOpen
  \bibfield  {author} {\bibinfo {author} {\bibfnamefont {C.~A.}\ \bibnamefont
  {Hidalgo}}, \bibinfo {author} {\bibfnamefont {E.}~\bibnamefont
  {Casta{\~{n}}er}}, \ and\ \bibinfo {author} {\bibfnamefont {A.}~\bibnamefont
  {Sevtsuk}},\ }\href {\doibase 10.1016/j.habitatint.2020.102205} {\bibfield
  {journal} {\bibinfo  {journal} {Habitat International}\ }\textbf {\bibinfo
  {volume} {106}},\ \bibinfo {pages} {102205} (\bibinfo {year}
  {2020})}\BibitemShut {NoStop}%
\bibitem [{\citenamefont {{OpenStreetMap contributors}}(2017)}]{OpenStreetMap}%
  \BibitemOpen
  \bibfield  {author} {\bibinfo {author} {\bibnamefont {{OpenStreetMap
  contributors}}},\ }\href@noop {} {\enquote {\bibinfo {title} {{Planet dump
  retrieved from https://planet.osm.org }},}\ }\bibinfo {howpublished} {\url{
  https://www.openstreetmap.org }} (\bibinfo {year} {2017})\BibitemShut
  {NoStop}%
\end{thebibliography}%


%

\end{document}